\documentclass{aastex631}

\newcommand\obs{$\theta_{\text{v}}$}
\newcommand\obsp{$\theta_{\text{v}}'$}
\newcommand\cp{$\theta_{\text{cp}}$}
\newcommand\cpp{$\theta_{\text{cp}}'$}
\newcommand\diff{$\theta_{\text{v}} - \theta_{\text{cp}}$}

\newcommand\ratio{$\frac{\theta_{\text{v}}}{\theta_{\text{cp}}}$}
\newcommand\ratiop{$\frac{\theta_{\text{v}}'}{\theta_{\text{cp}}'}$}
\newcommand\tp{$t_{\text{p}}$}
\newcommand\tpp{$t_{\text{p}}'$}
\newcommand\tobs{$t_{\text{obs}}$}
\newcommand\ujy{$\upmu$Jy}
\newcommand\kms{km\,s$^{-1}{\text{Mpc}^{-1}}$}
\newcommand\delra{$\delta_{\text{ra},i}$}
\newcommand\deldec{$\delta_{\text{dec},i}$}
\newcommand\ra{RA$_{\text{off}}$}
\newcommand\dec{Dec$_{\text{off}}$}

\usepackage{gensymb}
\usepackage{comment}
\usepackage{amsmath}
\usepackage{amssymb}
\usepackage{upgreek}
\usepackage{multirow}
\usepackage{makecell}
\setcellgapes{5pt}
\usepackage{hhline}

\begin{document}

\title{Revisiting GW170817 at milliarcsecond scale: high-precision constraints on jet geometry and $H_0$}

\author[0000-0002-0152-1129]{Kelly Gourdji}
\affiliation{CSIRO, Space and Astronomy, PO Box 76, Epping, NSW 1710, Australia}
\affiliation{Centre for Astrophysics and Supercomputing, Swinburne University of Technology, Hawthorn, VIC}
\affiliation{ARC Centre of Excellence for Gravitational Wave Discovery (OzGrav), Australia}

\author{Adam T. Deller}
\affiliation{Centre for Astrophysics and Supercomputing, Swinburne University of Technology, Hawthorn, VIC}
\affiliation{ARC Centre of Excellence for Gravitational Wave Discovery (OzGrav), Australia}

\author{Chris Flynn}
\affiliation{Centre for Astrophysics and Supercomputing, Swinburne University of Technology, Hawthorn, VIC}

\author{Taya Govreen-Segal}
\affiliation{School of Physics and Astronomy, Tel Aviv University, Tel Aviv 6997801, Israel}

\author{Cullan Howlett}
\affiliation{School of Mathematics and Physics, University of Queensland, Brisbane, QLD 4072, Australia}
\affiliation{ARC Centre of Excellence for Gravitational Wave Discovery (OzGrav), Australia}

\author{Kunal P. Mooley}
\affiliation{Indian Institute Of Technology Kanpur, Kanpur, Uttar Pradesh 208016, India}
\affiliation{Caltech, 1200 E. California Blvd.  MC 249-17, Pasadena, CA 91125, USA}

\author{Ehud Nakar}
\affiliation{School of Physics and Astronomy, Tel Aviv University, Tel Aviv 6997801, Israel}

\begin{abstract}
The historic detection of gravitational waves from the electromagnetically bright binary neutron star merger GW170817 enabled the first standard siren measurement of Hubble's constant ($H_0$). The accuracy and precision of this measurement depends crucially on how well the merger inclination angle is constrained, given its strong covariance with luminosity distance ($D_L$). Modeling the light-curve of the jet’s afterglow provides constraints on inclination, but is highly dependent on the similarly uncertain jet opening angle. Past studies have improved on this by invoking high-resolution radio observations, obtained through very long baseline interferometry (VLBI). We present a Bayesian visibility-plane model-fitting framework that provides a more informed and robust measurement of the viewing geometry of GW170817 and of $H_0$, by including all relevant VLBI data, robustly handling systematic uncertainties and rigorously sampling model parameter space.  By fitting new hydrodynamical afterglow models with a continuum of jet geometries, we obtain a viewing angle of $18\fdg3-20\fdg3$ (for a fixed cosmology with $D_L=40.7$\,Mpc, as used in most previous analyses). We extend our framework to fit for $D_L$ and $H_0$ directly, and marginalize over an ensemble of plausible peculiar velocity corrections to obtain viewing angle 16\fdg8$-$19\fdg2, $D_L=44.0\pm1.6$\,Mpc, and $H_0=65.5\pm4.4$\,\kms. Notably, the peak of our $H_0$ posterior is within $0.5\sigma$ of the early-Universe Planck $H_0$ value, but $1.7\sigma$ from the late-Universe SH0ES measurement. We discuss potential caveats and the implications of this result in the context of the current discrepancy between early and late-Universe measurements of the Hubble constant.

%We discuss potential caveats but, taken at face value, this result may challenge time-varying cosmology models.
%A larger ensemble of well-sampled afterglows is required to both further improve the statistical significance of the standard siren $H_0$ measurement and validate the jet models that underpin the inference. Our model-fitting framework can be applied to future VLBI datasets of gravitational wave events or gamma-ray bursts.

%The latter is consistent with, but more precise than, other works that have considered this important peculiar velocity systematic. 
\end{abstract}


\section{Introduction} \label{sec:intro}
Gravitational wave merger (GWM) events provide an independent avenue for measuring Hubble's constant \citep[$H_0$,][]{Schutz86,170817LVKh0}.  In the standard siren method, gravitational wave (GW) data from inspiralling or coalescing compact objects can be analyzed strictly through general relativity (GR) to measure the luminosity distance ($D_L$) to the merger event, independent of traditional distance ladders \citep{Holz2005,Nissanke2010}. However, the source redshift is not directly encoded in the GW signal. In the absence of an electromagnetic counterpart to identify the host galaxy, the redshift must be inferred statistically from galaxies within the GW localization volume (so-called `dark sirens'). In the case of `bright sirens', where an electromagnetic counterpart uniquely identifies the host galaxy of the merger, the host galaxy's recessional velocity ($V_H$, the velocity due solely to the expansion of the Universe) can be combined with the GW-derived luminosity distance to obtain $H_0$. In the local Universe, where the cosmological redshift is $z_{\text{cosmo}}<<1$, this can be roughly approximated with  $V_H\approx cz_{\text{cosmo}} \approx H_0D_L$. Extracting the cosmological redshift from the \textit{observed} redshift requires motion in the local gravitational potential (``peculiar velocities'') to be accounted and removed. At distances $\lesssim40$\,Mpc, typical peculiar velocities are $\gtrsim10\%$ of the cosmological redshift \citep[e.g.][]{Carrick15}.
%The determination of $V_H$ requires an accounting of the background velocities at the GWM position, which is done by measuring local peculiar velocities\footnote{Typical peculiar velocity magnitudes at a distance of $\sim40$\,Mpc are approximately 10 per cent of $V_H$ \citep[e.g.][]{Carrick15}.} in order to maintain independence from any electromagnetic cosmic distance ladders. The prototypical example of a bright siren comes from the landmark detection of the multiwavelength counterpart to binary neutron star GWM event GW170817 \citep{Abbott2017MW,abbott17GW}. Its relative proximity enabled identification of the host galaxy \citep[NGC4993,][]{Abbott2017MW}, which was measured to be at a distance of $40.7\pm2.4$\,Mpc (obtained using the surface brightness fluctuation method \citealp{Cantiello18}). This provided the first and, as yet, only opportunity to measure $H_0$ through the bright siren method: $70.0^{+12.0}_{-8.0}$\,\kms \citep{170817LVKh0}. %The determination of $V_H$ requires an accounting of the background velocities at the GWM position, which is done by measuring local peculiar velocities in order to maintain independence from any electromagnetic cosmic distance ladders.
%Through this method, \citet{170817LVKh0} obtained .
%A dominant source of uncertainty in such GW standard siren $H_0$ measurements arises from the intrinsic GR degeneracy between the inclination of the orbital plane ($\imath$) and the luminosity distance of the binary system.
A second source of uncertainty, and one that does not reduce fractionally with increasing source distance, arises from the degeneracy between the inclination of the orbital angular moment vector to the line of sight ($\imath$) and the luminosity distance of the binary system. Specifically, $h \propto \frac{\text{cos}\imath}{D_L}$, where $h$ is the `strain' or amplitude of the GW signal, true for both polarizations when the inclination angle is $\lesssim 45\degree$. The inclination can be significantly constrained if the two GW polarizations are precisely measured. However, this is not possible for our only bright siren, GW170817, as the event was viewed face-on (which makes the two polarizations difficult to distinguish) and significantly detected by only the nearly co-aligned Hanford and Livingston Laser Interferometer Gravitational-Wave Observatory (LIGO) detectors \citep{170817LVKh0}.

Fortunately, observations of the synchrotron afterglow caused by the relativistic gamma-ray burst (GRB) jet that was launched by the binary neutron star merger \citep[][]{AbbottGRB,Goldstein17} can help to abate this degeneracy. This is because the axis of the jet is expected to be orthogonal to the orbital plane of the merging neutron stars. For instance, \cite{Guidorzi17} analyze light-curve data of GW170817 up to 40 days post-merger to probe the GRB jet geometry (and, by extension, $\text{cos}\imath$) and constrain the viewing angle to between 25\degree\ and 50\degree and $H_0=74.0^{+11.5}_{-7.5}$\,\kms. However, the precision and accuracy of $H_0$ measurements that is achievable using the unresolved afterglow light-curve is limited by the light-curve's own geometric degeneracies, as its evolution depends largely on the ratio between the viewing angle and opening angle of the GRB jet \citep{Ryan20,NakarPiran21}. Hence, such approaches typically adopt a fairly specific forward model for the jet structure and microphysics to report high-precision viewing angle and $H_0$ constraints within a restricted family of models \citep[e.g.][]{wang21}. \citet[][hereafter referred to as M18]{Mooley18} use multi-epoch very long baseline interferometric (VLBI) observations to identify super-luminal proper motion of the merger's relativistic outflow. The proper motion provided largely orthogonal constraints to the copious multiwavelength synchrotron afterglow light-curve data (\citealp{NakarPiran21}; proper motion traces the difference between the viewing angle and jet opening angle, as will be detailed in Section \ref{methods}). M18 compare their multi-epoch VLBI data to hydrodynamical jet simulations to constrain the viewing angle of the jet of GW170817 to be between 14\degree and 28\degree. \citet{Ghirlanda19} use an additional VLBI data-set with higher spatial resolution, to constrain the size of the afterglow ($<2.5$\,mas at 207 days post-merger) and to estimate a viewing angle of $15$\degree$^{+1.0}_{-1.5}$. \citet{hotokezaka19} incorporated the viewing angle constraint from the M18 proper motion result into the GW bright siren measurement of $H_0$ to roughly double the precision to $68.9^{+4.7}_{-4.6}$\,\kms. \citet{dietrich20} introduce kilonova modeling as an additional inclination angle constraint and report $H_0=66.2^{+4.4}_{-4.2}$\,\kms, but the constraint from the kilonova is subdominant to the VLBI contribution. \cite{Mooley22} added jet proper motion measurements to the existing VLBI measurements of GW170817, leveraging the Hubble Space Telescope detection of the kilonova (which provides the position of the merger), thereby obtaining a viewing angle of 21\fdg9$_{-2.9}^{+3.3}$ and $H_0=71.5 \pm 4.6$\,\kms. More recently, \citet{palmese24} consider 3.5-years of unresolved afterglow light-curve data and a jet model with some priors informed by the values of jet opening angle inferred by the centroid motion analysis from \citet{Mooley22} to find $30\fdg4^{+2.9}_{-1.7}$ and $H_0=75.46^{+5.34}
_{-5.39}$\kms. This viewing angle estimate is significantly larger than the range inferred from the VLBI centroid-motion analyses. The tension highlights the sensitivity of viewing-angle constraints derived from unresolved afterglow light-curve modeling to assumptions about jet structure and microphysical parameters. \citet{Gianfagna24} provide a comprehensive assessment of GW170817 $H_0$ studies and show that the contrasting results between afterglow light-curve and the centroid motion approaches arise from the treatment of the late-time afterglow data, which appear to exhibit excess flux relative to jet models \citep{Hajela22}.

Such independent methods of measuring $H_0$ are highly prized given the tension \citep{h0tension} that exists between two very precise and independent early and late Universe measurements by the Planck Collaboration \citep[][$67.4\pm0.5$\,\kms]{planck20} and the Supernovae H0 for the Equation of State team \citep[SH0ES,][$73.04\pm1.04$\,\kms]{reiss22}, respectively. In this context, the value of an individual bright siren measurement will ultimately depend on its precision (and accuracy, of course). Indeed, the latest $H_0$ measurement to include new centroid motion information \citep{Mooley22} falls neatly between (and consistent with) those two independent measurements that are in tension.

Our study is concerned with obtaining the most informed constraint possible on the viewing angle of GW170817 to measure $H_0$. 
We improve on previous studies that invoke VLBI data in $H_0$ measurements \citep[e.g.][]{hotokezaka19,HowlettDavis20,Mukherjee21} and those that consider only one fixed peculiar velocity correction \citep[e.g.][]{Mooley22}, by including all astrometric information currently available and by considering multiple possible corrections. 
In the literature, either semi-analytically-produced synthetic jet afterglow models have been used (at the expense of model fidelity) to fit the afterglow data of GW170817, or sampling of only a limited part of the jet parameter space has been undertaken. Additionally, previous studies have considered only the centroid afterglow positions (as measured in the VLBI images) for fitting, thereby forgoing all other spatial information such as the size of the afterglow.
% and/or the uncertainties on the fitted parameters were not handled self-consistently. 
% Previous studies that invoke VLBI data in $H_0$ measurements either have not included all astrometric information presently available \citep[as is the case for][]{hotokezaka19,HowlettDavis20} or otherwise only consider one of the many plausible peculiar velocity corrections \citep[such is the case for][]{Mooley22}. 
% To facilitate model fitting, all previous analyses either used synthetic jet afterglow models that were produced semi-analytically (at the expense of model fidelity), or only sampled a limited part of jet parameter space. Additionally, in all cases, only the centroid afterglow positions as measured in the VLBI images were used in the fitting, thereby forgoing all other potential spatial information such as the size of the afterglow, and/or the uncertainties on the fitted parameters were not handled self-consistently. 
In this study, we address all of these limitations and make use of improved models that are based on sophisticated 2-D hydrodynamic simulations \citep{govreen-segal23}, to report our best estimate of the viewing angle of GW170817 and associated bright siren measurement of $H_0$. This methodology is described in Section \ref{methods}. The results that we obtain by applying our methodology, including viewing geometry constraints and measurement of $H_0$, are reported in Section \ref{results}. We discuss the implications and contextualize our results in Section \ref{disc}, before concluding in Section \ref{Conclusion}.

\section{ Data and Methods}
\label{methods}

Our study involves the analysis of all four existing datasets that measure the position of GW170817 at milliarcsecond scale. This includes two High Sensitivity Array VLBI epochs reported in \citet{Mooley18} (described in Section \ref{meth:hsa}), one Global-VLBI epoch presented in \citet{Ghirlanda19} (described in Section \ref{meth:global}) and a \textit{Hubble} space telescope measurement of the kilonova (described in Section \ref{meth:hst}). We require a fitting framework that can provide a robust estimate of the parameters of interest while handling the non-parameterized nature of the model brightness distributions. This can be provided by Bayesian inference operating directly on the observed radio interferometer visibilities. 
% As described above, past efforts to constrain the jet viewing angle have either extracted simplified parameters from the individual observations \citep[e.g.][]{Mooley22} or did not self-consistently estimate the uncertainties on the fitted parameters (M18). 
A Bayesian approach to modeling visibilities has been considered in a more general context before by \citet{Perkins15}, however their method is not suited to our application as it is restricted to brightness distributions that can be represented in a simple parametric form. We employ the GPU-accelerated software package \textsc{GALARIO} \citep{galario} as the backbone of our fitting, which, for a given fit instance, converts an input model image into model visibilities by sampling the \textit{uv}-plane of the visibility data. We use the nested sampler \textsc{Dynesty} \citep{Dynesty} via the \textsc{Bilby} interface \citep{Bilby}.\footnote{We note that \textsc{Bilby} is used to take advantage of its user-friendly interfacing with \textsc{Dynesty}, and is not used for any GW data analysis. We use a random-walk sampling method (\texttt{r-walk}) with 100 walkers (60 walkers for the M18 models analyzed in Section \ref{M18} of the Appendix), 4096 live points, multiple bounding ellipsoids, $nact=2$ (meaning each MCMC chain is thinned by half of the average accepted number of jumps per chain) and a stopping criterion of either $dlogz=1$ or $dlogz=0.1$. For the evidence based model comparisons shown in Table \ref{tab:fits}, the smaller value is used. The larger value is however selected over smaller thresholds to limit runtime given the high dimensionality and number of live points in all other fits. We verify that our parameter inferences, including evidence based Bayesian Model Averaging, are not affected by this choice by performing tests with the tighter stopping criterion ($dlogz=0.1$) to find no statistically significant differences.} Several fit analyses are conducted in our study. At minimum, they all possess the following common free parameters (see Table \ref{tab:params}):
\begin{itemize}
    \item Position angle (PA), the orientation of the model images, East of North in degrees;
    \item \ra\ and \dec, the offsets in right ascension and declination of the model, in arcseconds, from a reference position. We choose the phase center of the High Sensitivity Array observations as the position from which offsets refer to;
    \item $F_{0}$, a global scale parameter that converts the hydrodynamical simulation intensities into units of flux (Jy) for comparison with VLBI observations. This effectively amalgamates several degenerate microphysics parameters into a single amplitude scaling which is simpler to infer;
    \item $\delta_{F,i}$, nuisance parameter accounting for the uncertainty on $F_0$ for each VLBI epoch, $i$, expressed as a fraction of $F_{0}$;
    \item $\delta_{\text{ra,}i}$ and $\delta_{\text{dec},i}$, a pair of nuisance parameters for each fitted VLBI epoch, $i$, to reflect the astrometric systematic uncertainties arising from the imperfectly calibrated data.
\end{itemize}
We describe the priors on each of these parameters as we introduce the different datasets used in the following subsections. The complete list of fit parameters and their prior distributions is shown in Table \ref{tab:params} of the Appendix.

Our Bayesian visibility model-fitting framework works as follows. For each epoch, the model image corresponding to the current parameters being evaluated is generated. The pixels contained in the model image are normalized by dividing by the sum of the image pixels, and then the normalized model image is multiplied by $F_0(1+\delta_{F,i})$ and passed to a call to \textsc{GALARIO}'s \texttt{chi2Image} function, which rotates a given image by PA before it is Fourier transformed and sampled according to the $uv$ data of the given VLBI epoch. The resulting model visibilities are then translated by (RA$_{\text{off}}+$\delra) and (Dec$_{\text{off}}+$ \deldec). The $\chi^2$ is finally computed according to Equation 10 of \citet{galario}. The process is repeated for all epochs being considered and the total log-likelihood of the current parameter sample is computed by summing the above $\chi^2$ values and multiplying the result by $-0.5$.

\subsection{Description of HSA visibility data}
\label{meth:hsa}
The analyses presented in this study involve fitting data collected by the High Sensitivity Array (HSA) and originally presented and described in M18\footnote{The calibrated visibilities are hosted on Zenodo \citep{zenodo}.}. In particular, we use data corresponding to their Epochs 3 and 4\footnote{As reported in M18, the observations corresponding to their Epochs 1 and 2 had technical issues that severely impacted sensitivity at those epochs.}, centered at 4.5\,GHz. Each epoch is a concatenation of 3 and 4 observations respectively, and their respective average times post-merger are 74.7 and 230.25 days (we round these to 75 and 230 through the text henceforth, for simplicity). We read the data, as calibrated and concatenated in M18, into the Astronomical Image Processing System \citep[\textsc{AIPS};][]{aips} and average in time across scans ($\sim5$ minutes) to reduce the number of visibilities to fit. The first epoch is averaged to a frequency resolution of 30.5\,MHz and the second epoch to 126.5\,MHz (a finer resolution is used in the first epoch, as it includes the Green Bank Telescope which requires frequency dependent leakage corrections). As input, \textsc{GALARIO} requires visibilities as a simple flattened list of five data values: visibility real component, visibility imaginary component, visibility weight, $u$ coordinate, and $v$ coordinate. We create this list using \textsc{parseltongue}, which is a python interface for accessing \textsc{AIPS} data \citep{parseltongue}. Visibilities for which either (or both) polarizations have been flagged are excluded, and Stokes I values of the remaining visibilities are calculated by dividing the sum of the LL and RR values by two.

For the prior range on the corresponding astrometric uncertainty nuisance parameters (\delra\ and \deldec) we use the astrometric uncertainties reported in M18 on the compact check source, NGC 4993, located in the field. Across 7 HSA observations (our first HSA epoch consists of the concatenation of 3 observations, and the second consists of the concatenation of 4 observations), the root-mean-square variation on the position of the source is 0.14\,mas in RA and 0.49\,mas in Dec. These values are used to construct Gaussian priors for \delra\ and \deldec.  We assume a ten per cent fractional error on the flux scale of each VLBI dataset (i.e. a Gaussian prior distribution for $\delta_{F}$ centered on zero with $\sigma=0.1$). This was estimated in M18 by comparing VLBI fluxes to those measured using VLA-only interferometric data. M18 report peak flux densities of 58\,\ujy\ and 48\,\ujy\ for days 75 and 230, respectively.

\subsection{Inclusion of Global-VLBI data}
\label{meth:global}
The longer baselines, particularly in the North-South direction, of the Global-VLBI observations presented in \citet[][hereafter referred to as G19]{Ghirlanda19} provide higher resolution than the HSA data (the beam along the North-South direction is roughly 3 times smaller). Hence, inclusion of this data increases the constraining power on the size of the radio afterglow and potentially, by extension, the ratio between the jet viewing and opening angles. Importantly, these observations also provide a valuable additional astrometric data point 207 days post-merger \citep{govreen-segal23}, near the peak of the afterglow's light-curve ($\sim155$ days post-merger, \citealp{Makhathini21}). In order to include the global-VLBI  visibility data into our fit, we performed an independent reduction of this dataset, calibrating it using the same model for the frame-defining source J1312--2350 as was used in the HSA data reduction\footnote{The calibrated visibilities are hosted on Zenodo \citep{zenodo}.}. A full description of this reduction is provided in Section \ref{reduction} of the Appendix.  Given the comparable baseline lengths, observing frequency and calibrator, we apply the same uncertainties that we use for the HSA data so that $\delta_{\text{ra,3}}=0.14$\,mas and $\delta_{\text{dec,3}}=0.49$\,mas. We measure a peak flux density of 47\ujy\ and also similarly assume a ten per cent fractional uncertainty on the flux scale. We confirm that this is reasonable by comparing the measured flux to expectations based on the VLA flux measurement from M18 and the well-modeled temporal evolution of the light-curve around that time \citep[e.g.][]{Makhathini21}.

\subsection{Inclusion of HST astrometry in the optical}
\label{meth:hst}
We include a fourth astrometry epoch into the fitting framework, corresponding to the HST observation of GW170817's kilonova emission 8 days post-merger.  We use the astrometrically precise measurements derived in \citet[hereafter referred to as M22]{Mooley22}, which are in the GAIA/ICRF3 reference frame.  We apply the astrometric frame shifts derived in M22 (0.14\,mas in RA and 0.21\,mas in Dec) to place the day-8 position into the frame of the HSA observations (which are tied to their calibrator sources). As described in M22, we take the day-8 HST position of the kilonova to be the location of the merger itself (i.e. day-0). Therefore, we are anchoring the reference position offset of the models to the HST day-0 position. This is done by putting an informative prior on \ra\ and \dec. The peak of this prior is set to the position that is obtained when translating the HST best-fit position from the Gaia/ICRF3 frame into the frame defined by the assumed position of the calibrator of the HSA observations. The width of the Gaussian prior is the uncertainty of this frame translation (0.18\,mas and 0.34\,mas in RA and Dec, respectively; \citealp{Mooley22}). In likelihood evaluations that include HST astrometry, the drawn reference position offset (\ra, \dec) is compared to the HST measurement ($\text{RA}_{\text{hst}}$,$\text{Dec}_{\text{hst}}$) through Gaussian $\chi^2$ terms $\frac{(\text{\ra}-\text{RA}_{\text{hst}})^2}{\sigma^2_{\text{hst,ra}}}$ and $\frac{(\text{\dec}-\text{Dec}_{\text{hst}})^2}{\sigma^2_{\text{hst,dec}}}$ , where $\sigma_{\text{hst,ra}} =0.32$\,mas and $\sigma_{\text{hst,dec}}=0.19$\,mas are the statistical uncertainties on the HST measurement.

\subsection{New hydro-dynamical model images and scaling formalism}
\label{meth:scaling}
%To validate and assess the usefulness of the model fitting framework above, we first considered the set of five model afterglow images that were originally presented and analyzed in M18, before moving on to more sophisticated models described in this subsection. This re-analysis produced results that were generally consistent within the restricted range of parameter space sampled by the brute-force approach in M18. A complete description of the re-analysis and its results is provided in Section \ref{res:free} of the Appendix. 

%Having verified the fits to the 5 models originally presented in M18 using our framework, the aim of fully exploring the jet afterglow model parameter space can now be undertaken.
We wish to fully explore the jet afterglow model parameter space. To achieve this, one could consider producing a large set of models with a variety of jet types and different combinations of jet viewing and opening angles. This approach is computationally expensive and still would only finitely sample the large volume of plausible model parameter space. Instead, we apply a model scaling strategy, outlined below, to a new set of off-axis short gamma-ray burst (SGRB) jet afterglow models that are described in detail in \citet{govreen-segal23}. The models are based on full 2-D relativistic hydrodynamic simulations and their accuracy is tested and discussed in the aforementioned reference. The models are generated under the assumption that the jet is ultra-relativistic during the light-curve peak (true when viewing angle \obs\ is approximately $\theta_v<0.5$  rad, as in GW170817).  Each model-family is simulated with a given jet energy distribution (top hat or power-law with index $E_{iso}(\theta>\theta_c) \propto \theta^{-b} $ with $b=$3, 4, and 6, where \cp\ is the opening angle of the jet core at the peak of the light-curve) and initial jet core opening angle.

The following quantities from each hydrodynamic simulation are sampled across a range of times and jet viewing angles (\obsp) to construct two-dimensional model images: pixel intensity in arbitrary units, and pixel $x$ and $y$ coordinate values in cm. The following necessary metadata is also recorded for each observing angle model subset: the time in days post-merger at which the afterglow light-curve peaks (\tpp), the opening angle of the jet core at \tpp\ (\cpp), and the upper and lower time limits, in units of \tpp, between which the model can be trusted for scaling.  The latter upper and lower time limits define the observer-time interval over which the emission contributing significantly to the afterglow flux originates from shock radii that are included in the hydrodynamic simulation. This requirement arises because photons emitted at larger angles relative to the line of sight reach the observer at later times and therefore correspond to emission produced at smaller shock radii. The model images are invariant for jets with the same \ratiop\ and for observer time that is measured in units of \tpp. Hence, a given jet model family can be scaled from the `primed' parameter values described above, to some desired combination of time post-merger ($t_{\text{obs}}$ in days), viewing angle (\obs), opening angle (\cp), and observed afterglow light-curve peak time \tp\ as follows. First, the ratio between \obs\ and \cp\ and similarly between \obsp\ and \cpp\ of the viewing angle model subsets are computed. The model subset with \ratiop\ nearest to \ratio\ is selected for rescaling. The two model samples closest in time to $\frac{t_{\text{obs}}t_\text{p}'}{t_\text{p}}$ are linearly interpolated in both pixel coordinates and intensity values to obtain an interpolated model image. The interpolated model image pixel lengths are then scaled by the following expression:
\begin{equation}
\frac{t_\text{p}}{t_\text{p}'} \times \frac{(\theta_\text{v}' - \theta_{\text{cp}}')}{(\theta_\text{v} - \theta_{\text{cp}})}.
\label{eq:scaling}
\end{equation}
%where $t_p$ is the time in days at which the afterglow light-curve of the model being scaled peaks, and $t_p'$ is 155 days, the time post-merger at which the GW170817 afterglow light-curve peaked.
Lastly, the pixel coordinates are converted from cm to angular units using a given source distance $D_L$. Where we do not marginalize over $D_L$ in our fits (see next section), we have fixed $D_L$ to 40.7\,Mpc, the best-fit luminosity distance of NGC4993, the host galaxy of GW170817, obtained via the surface brightness fluctuation method \citep{Cantiello18}. The latter analysis is done to enable direct comparison to geometry results reported in the literature, which widely use a fixed cosmology with $D_L=40.7$\,Mpc.

We fold this model scaling formalism into the fitting framework by introducing two new parameters, (\diff)\ and \ratio, with uninformative priors, unless stated otherwise. These two parameters are used to generate a model image for each VLBI epoch in a given likelihood evaluation. To first order, \diff\ controls the amount of proper motion and \ratio\ controls the size of the afterglow. The image resolution is decreased from the native 400 by 400 pixels to 80 by 80 pixels, which corresponds to a pixel resolution of order 100 microarcseconds, assuming $D=40.7$\,Mpc. The observed peak time of the afterglow light-curve \tp\ is fixed to 155 days \citep{Makhathini21} unless otherwise indicated.

%, except where a prior on \ratio\ informed by When incorporating information from the afterglow light-curve of GW170817, we introduce \tp\ as a fit-parameter (as opposed to fixing it to 155 days) and use informative Gaussian priors for both \tp\ and \ratio\ based on the posteriors that resulted from the comprehensive GW170817 lightcurve analysis published in \citet{Makhathini21}.
%as described in \ref{meth:fit1} and \ref{meth:hst}.The likelihood evaluation is otherwise carried out in the same way.

%This scaling formalism is valid so long as we are probing  the ultra relativistic regime.  

%\citet{NakarPiran21} demonstrate the inability to individually discern \obs\ and \cp\ using the afterglow light-curve alone. They show that the width of the afterglow light-curve provides a constraint only on \ratio  \citet{govreen-segal23}. Additional information is required to break this degeneracy. Talk about centroid and size of afterglow but also late time detections can apparently enable measurement of $\Gamma$. show that measurement of the afterglow flux centroid constrains \diff.

\subsection{Fitting for Hubble's constant}
\label{meth:h0}
The final tier of our fitting framework incorporates relevant cosmology and GW information and parameters. We largely follow the strategy presented in \citet[]{HowlettDavis20}, where the total and peculiar velocities of various plausible galaxy groups containing GW170817's host (NGC~4993) are considered. Three new fit parameters are introduced: the Hubble Constant ($H_0$), the luminosity distance $D_L$, and the redshift of the galaxy  hosting NGC4993, which we multiply by the speed of light ($cz$). The prior distributions of $D_L$ and $H_0$ follow a power-law distribution with index 2 and $-1$, respectively.  Three new likelihoods are added in log-space to those outlined in the previous subsections:
\begin{itemize}
    \item $\mathcal{L}(\text{GW}|D_L,\text{cos}\imath)$, where cos$\imath$ is the inclination of the GW event and is equivalent to cos($\pi-\theta_{\text{v}}$). To compute the likelihood, we create a kernel density estimator (KDE) using the posterior samples from \citet[see Figure \ref{fig:kde} in the Appendix]{170817props};\footnote{\url{https://dcc.ligo.org/LIGO-P1800061/public}}

    \item  observed heliocentric  velocity ($cz$). The likelihood is evaluated by computing $\chi^2$ of a value drawn from a flat prior to an empirical catalog value and its associated uncertainty. We consider the catalogue values reported in Table 2 of \citet[]{HowlettDavis20};\footnote{We exclude the outlier that is  955 on the basis that NGC~4993 is only marginally associated with it (see e.g. \citealp{Hjorth17})}

    \item $\mathcal{L}(v_p|D_L,cz,H_0)$, where $v_p$ is the peculiar velocity of the galaxy group containing NGC~4993. $v_p$ can be recovered from $z_\text{p}$, the peculiar redshift, which is defined as:
    \begin{equation}
        1+z_\text{p} = \frac{1+z_\text{cmb}}{1+z_\text{cosmo}(cz,D_L,H_0)}.
    \end{equation}
    Here, $z_\text{cosmo}$ is the redshift due purely to the expansion of the Universe, and $z_\text{cmb}$ is obtained by converting the heliocentric observed  redshift into the CMB frame, which corrects for the Solar system’s peculiar motion with respect to the CMB dipole. The likelihood of $v_\text{p}$ is calculated by comparing it to an empirical value and its uncertainties determined using \textsc{pvhub}\footnote{\url{https://github.com/KSaid-1/pvhub}} and some chosen peculiar velocity reconstruction, which is effectively a model choice as discussed below.
\end{itemize}

Because the latter two likelihoods depend on the choice of  catalog and peculiar velocity reconstruction, we consider permutations of the four reconstructions available in \textsc{pvhub}\footnote{2MPP-SDSS \citep{Said20,Peterson22,Carr22}; 2MPP-SDSS\_6dF \citep{Said20}; 2MRS \citep{2MRS}; 2MPP \citep{Carrick15}.} and the 7  redshift catalogs from \citet[][see Table 2]{HowlettDavis20} and choose to report the resulting Bayesian model averaged $H_0$ value in Section \ref{results}, where the posterior samples are combined and weighted by each model's Bayes evidence. We use an uncertainty for the peculiar velocity correction of 150\,\kms, which has been widely used in the literature for nearby galaxies in low-redshift cosmological analyses \citep{Carrick15,170817LVKh0}.
%, such that all other combinations result in values that fall between the reported extrema. 

%$p(D_L)$ is proportional to $D_L^2$ and $p(H_0)$ to $H_0^{-1}$. 

\section{Analysis and results}
\label{results}
%To validate and assess the usefulness of the model fitting framework above, we first consider the set of five model afterglow images that were originally presented and analyzed in M18 in Section \ref{M18}, before moving on to more sophisticated models described in Section \ref{meth:scaling} and used in Sections \ref{res:scaling} and \ref{res:h0}.

To validate and assess the usefulness of the model fitting framework above, we first considered the set of five model afterglow images that were originally presented and analyzed in M18, before moving on to more sophisticated models described in Section \ref{meth:scaling}. This re-analysis produced results that were generally consistent within the restricted range of parameter space sampled by the brute-force approach in M18. %A complete description of the re-analysis and its results is provided in Section \ref{res:free} of the Appendix. 
A full account of the results from the re-analysis of M18 is presented in Section \ref{res:free} of the Appendix. In brief, the re-analysis of M18 (i.e. fitting 5 discrete models to only the days 75 and 230 HSA VLBI data) using our framework produces statistically robust results that are mostly consistent with M18's original goodness of fit findings. However, when we repeat the analysis but include the optical HST astrometry (Section \ref{meth:hst}), we find that \textit{none} of the models fit the joint HSA-HST dataset well (see Fig. \ref{fig:hstfit}), indicating that they do not adequately sample the full plausible range of jet parameter space. We consider this to be further motivation for the analyses that are performed in this section.  

\subsection{Direct fits to jet geometry}
\label{res:scaling}
We adopt the methodology presented in Section \ref{meth:scaling} to continuously explore the full plausible physical parameter space of jet viewing and opening angles, for four different jet-types of the model presented in \citet{govreen-segal23} (one top-hat and 3 different power-law energy distributions). We begin by fitting the HSA-HST dataset (i.e. days 0, 75 and 230). According to the model scaling relation used (Equation \ref{eq:scaling}), \diff\ is being constrained directly, as is reflected by the resulting narrow posteriors across all four fits (Table \ref{tab:fits} of the Appendix). The resulting log evidences of the 4 types of jet models do not indicate any significantly preferred models (Table \ref{tab:fits}) following \citet{Jeffreys61}, and the posteriors of the  \diff\ parameter are fairly consistent across the models, all falling in the range 13\degree\ to 16\degree (68\% credible level). This is consistent with expectations from the analytical model of the jets, which show that near and after the light-curve peak (\tobs\ $>0.5$\tp) the emission becomes less sensitive to the angular structure of the jet, as it becomes dominated by jet core material close to the observer's line of sight \citep{govreen-segal23}. This behavior is most robust for observers located a few jet core angles off-axis (\ratio\ $\gtrsim2$), as is relevant here. Notably, the first VLBI epoch is at approximately $0.5$\tp.
For the remainder of our analysis, we consider the power-law jet model-family ($b=6$), though we emphasize that this choice should have negligible relative effect for the reasons explained above.

% We find strong consistency between the resulting posterior distributions of \diff\ for all four jet-types that were considered.
%The right hand sub-figure displays the posteriors of \diff\ and \ratio in 2-D space for PL($b=6$), as an example.

\begin{figure}[h]
%\plotone{plot_4_corners.pdf}
\centering
\includegraphics[width=\textwidth]{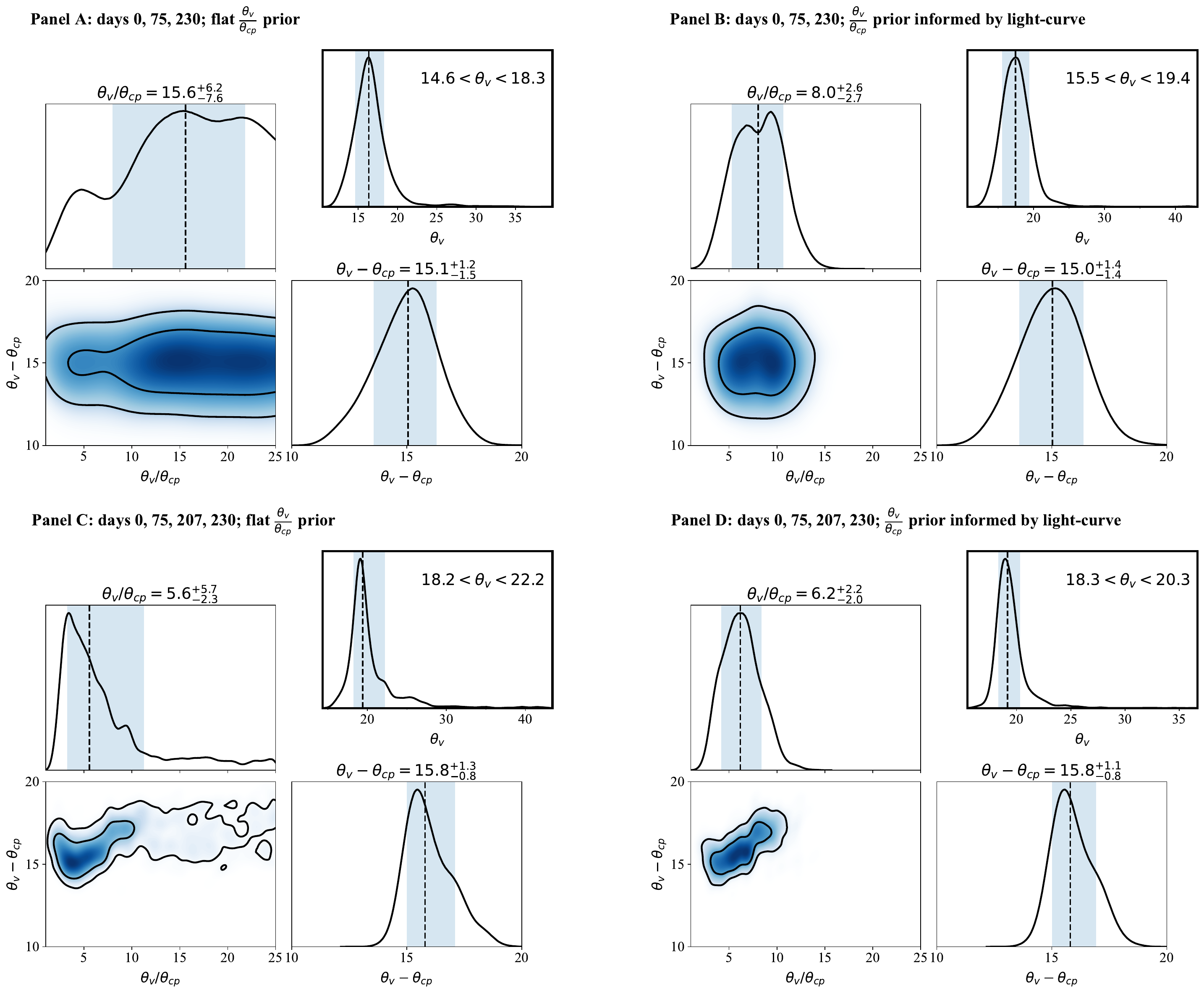}

\caption{Corner plots of the \diff\ (in degrees) and \ratio\ posteriors resulting from fits that include (panels A to D): A) the HSA and HST data (days 0, 75, 230), B) same but with \ratio\ priors informed by the afterglow light-curve, C) all astrometric data (days 0, 75, 207, 230), D) same but with \ratio\ priors informed by the afterglow light-curve. The corresponding viewing angle (\obs) posterior samples, recovered algebraically by rearranging the \ratio\ and \diff\ posterior samples, are represented by a histogram in the top right corner of each sub-figure. The posterior median and 68\% credible interval are reported and denoted by vertical dashed lines and shaded regions in the histograms, respectively. The corner plot contours denote the 68\% and 95\% credible levels. All four fits included here used the power-law ($b=6$) jet model-family.  These posterior distributions are visualized using \textsc{GetDist}, which employs adaptive kernel density estimation to generate smooth approximations to the underlying distributions. These smoothed curves are used exclusively for visualization purposes. All quoted parameter estimates and credible intervals are computed directly from the posterior samples.}
\label{fig:geometries}
\end{figure}

\begin{figure}
    \centering
    %\plottwo{model_scaling_angle_diffs_posterior.pdf}{plb6_corner.pdf}
    %\textit{Top left}: \diff\ posteriors for the four jet-types considered. \textit{Top right}: Example posteriors of \diff\ and \ratio\  for the PL ($b=6$) jet model.  \textit{Bottom}:
    \includegraphics[width=0.9\linewidth]{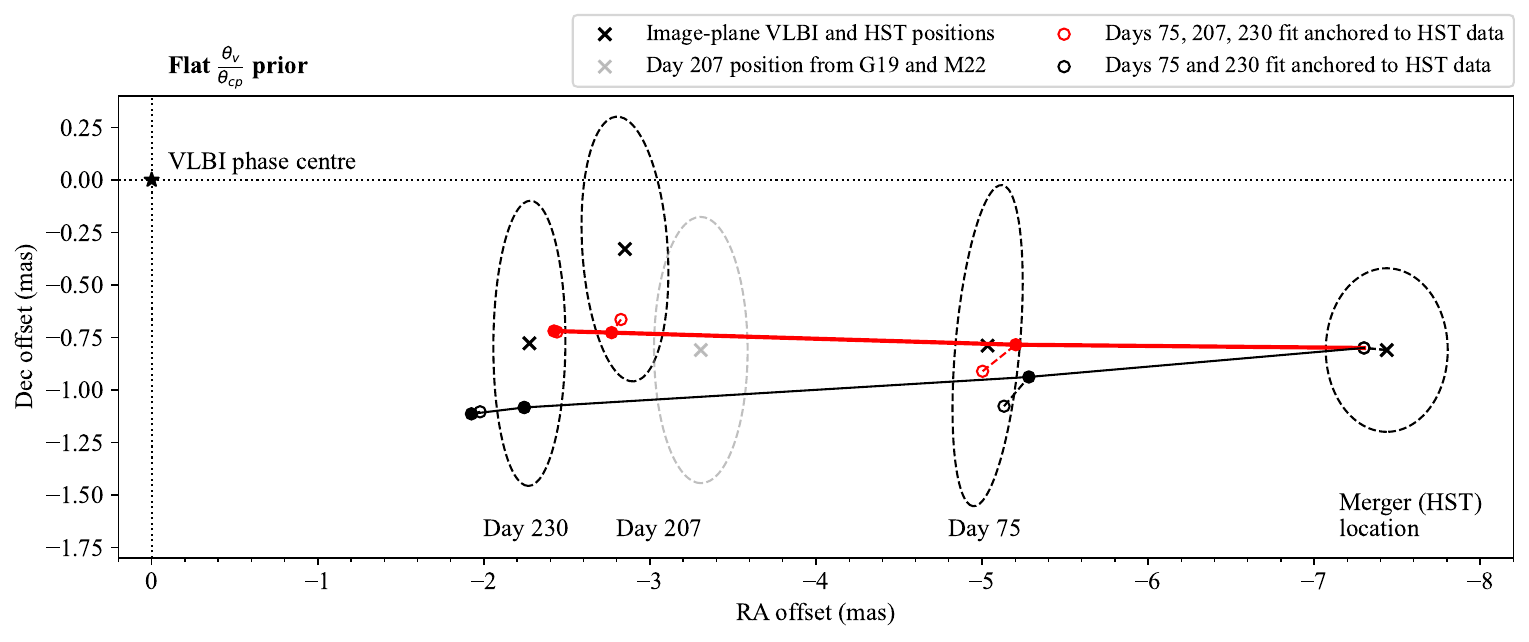}
    \includegraphics[width=0.9\linewidth]{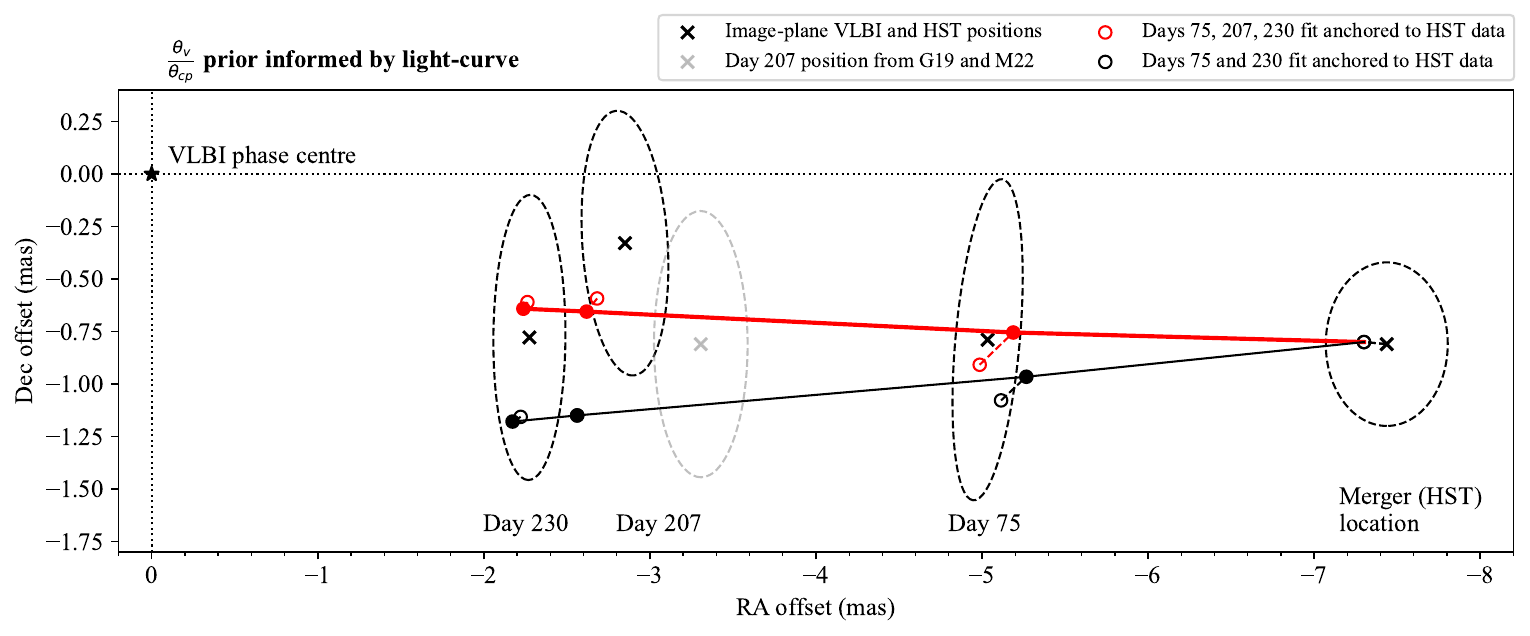}
    \caption{Astrometry plots showing the model-scaling fit results, with uninformative priors on jet parameters in the top panel, and priors informed by light curve information in the bottom panel. Two different fits are shown in each: day-207 data omitted (black) and included (red). The closed circles are the flux centroids of a representative afterglow model resulting from the fit, for each epoch, and are anchored to the fitted location of the merger (right-most open black circle, i.e. \ra\ and \dec). Since \ratio\ is unconstrained in the black fit of the top-panel, we set it to 15. For the `black' fits, we have included where the flux centroid at day 207 would be for that model for reference. The solid circles may not be perfectly aligned, given that the hydrodynamical model images used vary in pixel size between epochs. The open circles represent the astrometric nuisance parameters and indicate where the centroid would appear on the sky after including the systematic error inferred for the data on that epoch (\delra\ and \deldec).  The black X-shaped markers correspond to the image-plane positions of GW170817 from the VLBI data and the HST kilonova position (day 0, i.e. merger location). The $1\sigma$ uncertainty ellipse on each of these are determined using the values from Table \ref{tab:astrometry}, where the systematic error ellipse has been convolved with the statistical errors from the image-plane fit along the beam. The grayed ellipse represents the image-plane position of the afterglow 207 days post-merger that was obtained in G19 and used in M22. 
    %(c.f. the black data-point -- we have re-reduced the G19 data for our analysis, which resulted in a slightly different position).
    All data are in the HSA frame and are relative
to the HSA observations’ phase center (denoted by a star).}
    %For simplicity, an infinitesimal pixel size has been used for the right-most filled point, which denotes the model merger position before propagating systematic astrometric nuisance parameters.
    %The uncertainty ellipse on each of these are determined using the values from Table \ref{tab:astrometry}, where the systematic error ellipse (along RA and Dec) has been convolved with the statistical errors from the image-plane fit along the beam.
    \label{fig:scaling}
\end{figure}

In contrast to \diff, we have little constraining power on \ratio, as demonstrated by the corner plot in Panel A of Figure \ref{fig:geometries}.  This is because the \ratio\ parameter affects the size of the afterglow more than it does the proper motion, and the resolving power of the HSA VLBI data is comparable in scale to the models that have the  smallest ratios (i.e. largest afterglow widths) explored.
%(can use example from b4 where smaller ratio 5 gave more consistent proper motion but was disfavored in the fit because it gets too big).
Furthermore,
since \obs$-$\cp $\approx$ \obs\ for \cp $\ll$\obs, increasing the ratio beyond 10 has little effect on the resulting viewing angle\footnote{viewing angle increases non-linearly with decreasing angle ratio. With some alegbraic rearrangement of the angle difference and angle ratio parameters one obtains: \obs$\propto(\frac{\text{ratio}}{\text{ratio}-1})$}.
%given that \obs$\propto(\frac{\text{ratio}}{\text{ratio}-1})$ (obtained after some algebraic rearrangement of the angle difference and angle ratio parameters), increasing the ratio above $\sim10$ has little relative effect on the resulting viewing angle (i.e. viewing angle increases non-linearly with decreasing angle ratio).
%Hence, we can say viewing angles less than approximately $20\degree$ are preferred.
We present results of the fit in Figure \ref{fig:scaling} using black data points, where we have set \ratio$=15$. The fit displays much better agreement with the image-plane centroid positions of the afterglow relative to the M18 models (Figure \ref{fig:hstfit} in the Appendix), particularly when including the astrometric nuisance parameters which act to pull the models in a direction that lowers the $\chi^2$ of the fit to the visibility data. We emphasize that the image-plane centroid positions (black crosses) should not be taken as the `true' or `best' positions of the afterglow; fundamentally, fitting directly in the visibility plane is more accurate, as it is independent of image de-convolution settings. Furthermore, the image-plane positions were obtained using a (potentially overly simplistic) singular circular Gaussian model component. Rather, the image-plane positions are included merely for reference and to serve as an approximate goodness of fit check.

While we have little constraining power on \ratio, the afterglow light-curve actually provides entirely orthogonal information, since the width of the light-curve peak (whilst in the ultra-relativistic regime) analytically probes \ratio\ \citep{NakarPiran21}.  \citet{Makhathini21} present a comprehensive analysis of the panchromatic afterglow light-curve between 0.5 and 940 days post-merger, and find $5<\frac{\theta_{\text{v}}}{\theta_{\text{cp}}}<10$ and $151<$\tp$<159$, with 68\% confidence. We use this information to introduce a Gaussian prior distribution for \ratio\ as well as a new fit parameter, \tp\ (previously fixed to 155 days), with a Gaussian prior. We repeat the fit with these modifications to find 15\fdg5$<$\obs$<$19\fdg4 (see Panel B of Figure \ref{fig:geometries}). 
%Combining this with our constraint on \diff, we find $14^{\circ}<\theta_{\text{v}}<20^{\circ}$ and $0.7^{\circ}<\theta_{\text{cp}}<1.8^{\circ}$. 

\begin{figure}[h]
\plotone{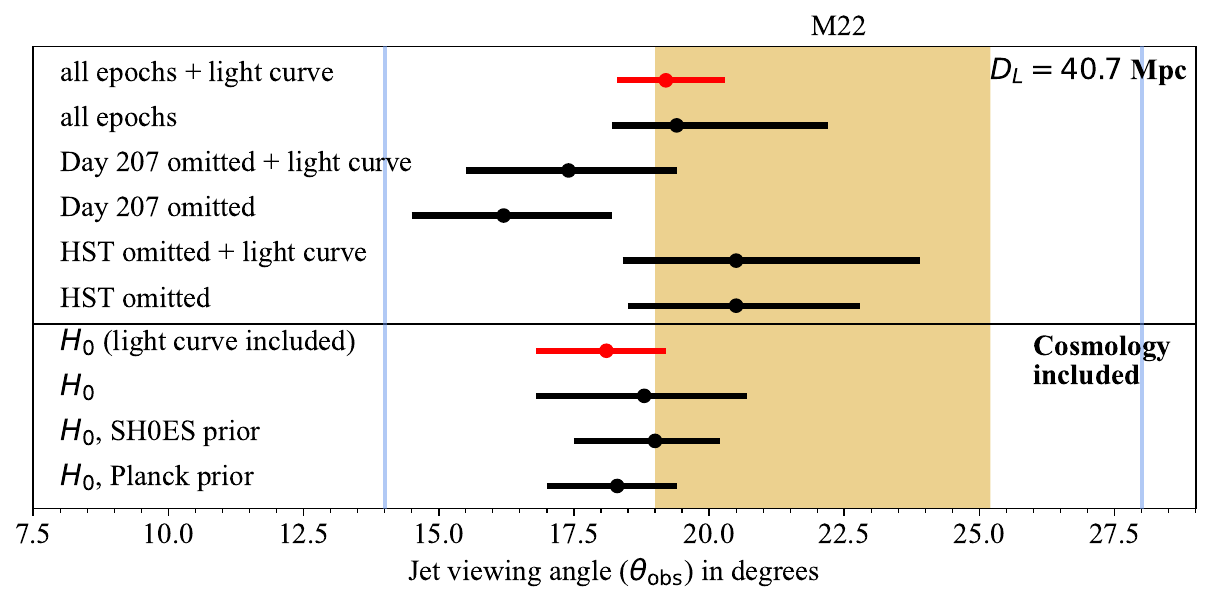}
%where instead of invoking BMA we simply report the extrema across all 28 peculiar velocity corrections considered to save compute time.
%\includegraphics[width=\textwidth]{M18_HST_astrometry.pdf}
\caption{Summary of 68\% credible intervals for the jet viewing angle from our various fits. The top panel excludes cosmology and GW information from the fits and the luminosity distance is fixed to $40.7$\,Mpc. The six cosmology-free fits shown correspond (in descending order) to including: all 4 astrometric datasets plus light-curve informed priors (our most informed and `best' fit of the 6, emphasized in red), all 4 astrometric datasets, day-207 data omitted but including light-curve informed priors, day-207 data omitted, day-0 omitted but including light-curve informed priors, and lastly day-0 omitted. The bottom panel shows results when including cosmology into the fit as explained in Section \ref{meth:h0}. The top `$H_0$' result refers to  the Bayesian Model Averaged viewing angle obtained using uninformative priors on $H_0$ across the 28 peculiar velocity corrections as described in Section \ref{res:h0}. The next one refers to the result obtained using flat priors on \ratio\ and fixing \tp\ to 155 days. The bottom two results correspond to having applied an $H_0$ prior distribution using the SH0ES result and Planck result, also as described in Section \ref{res:h0}. The 90\% credible interval on viewing angle obtained by \citet{Mooley22}, which represents the previous best estimate in the literature (for $D_L=40.7$\,Mpc), is denoted by a yellow band. The vertical blue lines bound the constraint on \obs\ reported in \citet{Mooley18} where light-curve information and only the two HSA astrometric datasets and a small selection of models were considered.}
\label{fig:obs}
\end{figure}

We now include a third VLBI data point in our Bayesian model-fit, at day-207 post-merger, as described in Section~\ref{meth:global}. This results in a far more useful constraint on \ratio\ of $5.6_{-2.3}^{+5.7}$, before even introducing information about the light-curve (Panel C of Figure \ref{fig:geometries}). This additional data point adds constraining power on the models via two primary avenues. First, the day-207 position relative to the day-75 and day-230 positions significantly constrains the deceleration of the afterglow centroid. Without the day-207 data, the systematic uncertainties at each epoch allow considerable variations and hence only a weak constraint on the deceleration. Second, the higher angular resolution of the day-207 observation in the North-South direction is better able to discriminate against models with very low angle ratios for which the afterglow size at day-207 is becoming increasingly large (though the low S/N of the data makes these constraints modest). Specifically, the best-fit model when considering only the HST, day 75 and day 230 data predicts a position angle, PA, of $91\degree$ East of North for the afterglow motion in the plane of the sky, which enables the model to best reproduce the observed motion in right ascension (where the positional precision is highest) from day 0 to 75 and day 75 to 230 (see Figure \ref{fig:scaling}). All three data points can be acceptably fit when incorporating the allowed systematic positional uncertainties. However, the day-207 data, for which the best fit position is slightly North of the days 75 and 230 data, is incompatible with that orientation, requiring a slightly smaller position angle. To further illustrate this, we have shown in Figure \ref{fig:scaling} where the day-207 flux centroid is expected to be in the fit that does \textit{not} include the day-207 data (black data-points). At the same viewing angle offset and ratio, however, a smaller rotation (i.e. simply rotating the black model data-points to the North) would still lead to too much displacement at day 230 (and far too much at day 207). A significant compensation in the viewing angle parameters is then required, with the net effect of reducing the predicted displacement at day 207 appreciably. This is best fit by a combination of increasing the viewing angle offset (\diff) (which reduces the displacement across all epochs) and decreasing \ratio. The latter increases the displacement between day 207 to 230, relative to that from day 75 to 207. Importantly, however, the displacement between day 75 and 207 suddenly becomes too large at ratios less than about 2.5 (a physical effect apparent also in \citealp{govreen-segal23}, see Figure 11 therein), thus significantly contributing to the lower constraint on \ratio visualized in Figure \ref{fig:geometries}.
%This is because the model now has to try to match the proper motion between days 75 and 207 post-merger. The observed image-plane displacement between these epochs is approximately 2.2 mas in RA. Displacements of that magnitude can only be achieved for models with smaller angle ratios and larger angle differences. However, angle ratios that are much smaller than 5 are now disfavored by the higher resolution global-VLBI data, as larger ratios blow-up the size of the radio afterglow \textbf{KG: this isn't true (a day 0+207 fit does not yield a lower limit). Still trying to understand why lower limit is obtained when we fit 4-epochs}. Similarly, the image-plane displacement between days 207 and 230 is about 0.6 mas, which can only be achieved with smaller angle ratios and, this time, smaller angle differences. 
%Overall, we find that it is challenging to simultaneously reproduce the observed displacements between days 75--207 and days 207--230 (see red data-points in the astrometry plot of Figure \ref{fig:scaling}).
%These displacements can be adequately fit by models with relatively low values of viewing angle (as this leads to relatively more motion earlier than later), but implies that one VLBI observation does exhibit a systematic astrometric error on the level of 1.5$\sigma$. 
We note that it would not have been possible to obtain an adequate fit with the day-207 position presented in G19 and used by M22 (background gray ellipse in Figure \ref{fig:scaling}), illustrating the importance of carefully matching the reference frames used in different VLBI observations. The net effect of including the day-207 data into the fit is a slightly larger resulting angle difference (\diff$=16^{\circ+1}_{\,-1}$) and a more constrained angle ratio (see Figure \ref{fig:geometries}). This yields a 68\% credible interval on viewing angle of $18.2<$\obs$<22.2$. We caution that, where a flat \ratio\ prior is used, the associated implied prior on \obs\ acts to push the posterior to artificially smaller viewing angles. We provide an assessment of the implied \obs\ prior in section \ref{app:prior} of the Appendix.  Finally, as above, we fold in light-curve information by including an informative prior on \ratio\, and \tp\ to obtain $18\fdg3<$\obs$<20\fdg3$ and $2\fdg3<$\cp$<4\fdg7$. Our \obs\ results (for $D_L=40.7$\,Mpc) are summarized and compared to relevant results from the literature in Figure \ref{fig:obs}.

%Lastly, we include an informative prior on \ratio\, using a Gaussian distribution centered on 7.5 and $\sigma=2.5$, and also fit for the peak of the light-curve $t_p$, taking a Gaussian distribution centered on 155 and $\sigma=4$, all as reported in the comprehensive light-curve analysis published in \citet{Makhathini21}. We find $17.0<$\obs$<21.9$ (68\% confidence level). 

\subsection{Direct measurement of $H_0$}
\label{res:h0}
\begin{figure}[h]
%\plotone{H0_results_pub_no250.png}
\plotone{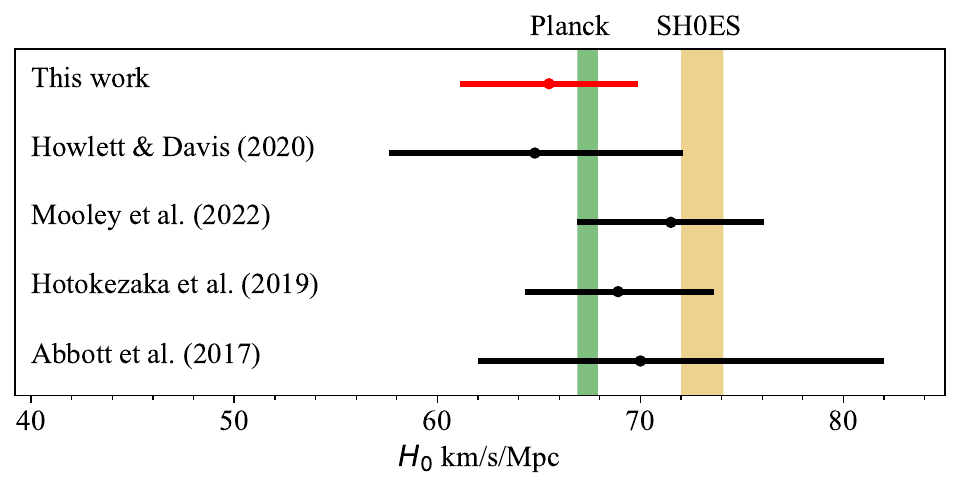}
%\includegraphics[width=\textwidth]{M18_HST_astrometry.pdf}
%\caption{The 68\% credible intervals of $H_0$ from our analysis compared to those from a selection of relevant previous works using GW170817. Our results are presented using a very conservative peculiar velocity uncertainty (250 \kms, red) and the widely used value of 150 \kms (black). We present 2 sets of results: one using  catalog 6 and peculiar velocity reconstruction TwoMPP, the other using  catalog 7 and reconsturction TwoMPP SDSS. These combinations produce the smallest and largest values of $H_0$, respectively, among all possible permutations that were considered in \texttt{pvhub}. In each set, results are presented for fits that 1) exclude the day 207 VLBI data, 2) include all 4 astrometric data sets, and 3) same as the latter but with priors on \ratio\ and \tp\ informed by analysis of the afterglow light-curve. The $1\sigma$ extent of two independent $H_0$ experiments, \citet{reiss22} and \citet{planck20}, that are in tension with one another are included for reference.}
\caption{The 68\% credible interval of our most informed measurement of $H_0$ ($65.5\pm4.4$\,\kms) compared to those from a selection of relevant previous works using GW170817, as discussed in Section \ref{disc:h0}. Our measurement is obtained by combining posterior samples across 28 combinations of galaxy group catalog and peculiar velocity reconstruction and weighting them by the Bayes evidence of their respective fit (i.e. Bayesian Model Averaging). The $1\sigma$ extent of two independent $H_0$ experiments, \citet{reiss22} and \citet{planck20}, that are in tension with one another are included for reference.}
%We present 2 sets of results: one using group catalog 6 and peculiar velocity reconstruction TwoMPP, the other using group catalog 7 and reconstruction TwoMPP SDSS. These combinations produce the smallest and largest values of $H_0$, respectively, among all possible permutations that were considered in \texttt{pvhub}. In each set, results are presented for fits that 1) exclude the day 207 VLBI data, 2) include all 4 astrometric data sets, and 3) same as the latter but with priors on \ratio\ and \tp\ informed by analysis of the afterglow light-curve.
\label{fig:H0}
\end{figure}

We apply the methodology outlined in Section \ref{meth:h0} and incorporate all available relevant information (days 0, 75, 207, 230 and \ratio\ prior informed by the light-curve) into the fit, obtaining $H_0=65.5\pm4.4$ \kms, which is the Bayesian model average value of all 28 peculiar velocity correction $H_0$ fits\footnote{The posterior samples are hosted on Zenodo \citep{zenodo}.}.  Our result is compared to other relevant results in the literature in Figure \ref{fig:H0} and they are contextualized in Section \ref{disc:h0}.  The largest natural logarithm of the Bayes factor across all 28 peculiar velocity correction $H_0$ fits is 1.8, which corresponds to marginal evidence in favor of galaxy  catalog 763 \citep[high density catalog from][]{crook2007,crook2008} and peculiar velocity reconstruction \texttt{TwoMRS\_redshift} over galaxy catalog 45466 (from \citealp{KourkchiTully17} but excluding those galaxies identified by \citealp{Hjorth17} as being only loosely associated with that group) and reconstruction \texttt{TwoMPP\_SDSS\_6dF} (see \textsc{pvhub} for details on the reconstructions and \citealp{HowlettDavis20} for details on the catalogs).  

To assess the impact of the light-curve information in our $H_0$ measurement,  we run three additional sets of fits with flat priors on \ratio\ and \tp\ set to 145, 155 and 165 days. Respectively, these yield $H_0=65.5\pm4.4$\kms, $H_0=65.8\pm4.5$\kms, and $H_0=66.3\pm4.6$\kms. 

The viewing angle results reported in the previous section were all obtained by assuming the surface brightness fluctuation luminosity distance to NGC~4993 ($D_L=40.7$\,Mpc) in our fits. Our $H_0$ fit results in $D_L=44.0\pm1.6$\,Mpc (this discrepancy is discussed in Section \ref{disc:view}). Having fit for $D_L$ directly, we can report the corresponding constraint on the viewing angle: 16\fdg8$<$\obs$<$19\fdg2 (\ratio\ and \tp\ priors informed by the light-curve) and 16\fdg8$<$\obs$<$20\fdg7 (uninformed \ratio\ prior and \tp$=155$ days). To further assess the sensitivity of \obs\ on the assumed distance/cosmology, we perform fits where the prior on $H_0$ is set to the posterior probability from \citet{reiss22} in one case, and to that of \citet{planck20} in another, to yield 17\fdg5$<$\obs$<$20\fdg2 and 17\fdg0$<$\obs$<$19\fdg4, respectively. 

%Impact on H0 of incorporating 4 versus 3 epochs. Seems incorporating 4 epochs systematically increases H0 by 1.5\%.

\section{Discussion}
\label{disc}
\subsection{Lessons learned for future events}
We assess the impact of probing GW170817 at sub-mas resolution at the four astrometric epochs available (days 0, 75, 207, and 230 post-merger), to help guide the VLBI observing strategy of future electromagnetically bright GW merger events. The fits from Sections \ref{res:free} and \ref{res:hst} that are summarized in Figure \ref{fig:hstfit}, illustrate the tremendous value of the precisely calibrated HST position of the kilonova from M22 that provides a day-0 anchor for the model fitting. Because the afterglow centroid motion is greatest at early times, the day-0 position provided by the kilonova provides even more constraining power on the position angle of the jet orientation than might be naively assumed based on the separation in time from the first VLBI epoch.
%The poor fits of the M18 models when the HST data is included  (shown in Figure \ref{fig:hstfit}) also demonstrate the critical importance of exploring the full plausible jet geometry parameter space (the result of which is shown in Figure \ref{fig:scaling}).
To assess the value of the day-0 point in more detail, we run a model interpolation fit but omit the HST data, allowing the reference position offset to be fit freely with flat priors. This returns a much larger value for \diff\ of $19\pm2$ (c.f. $16\degree\pm1\degree$), which in turn leads to a preference for large \ratio\ values to counter this effect. Since \obs$\propto(\frac{\text{ratio}}{\text{ratio}-1})$, it is difficult with this constraint on \diff\ to obtain viewing angles much less than $20$\degree. %(this would also start to require \cp\ to be less than the initial opening angle of the jet models).
We repeat this fit but include informative priors on \ratio\ and \tp\ based on the afterglow light-curve. This results in  \ratio$=9^{+2}_{-3}$ and angle difference value \diff\ of  $18.2^{+2.7}_{-2.2}$ degrees, which together yield 18\fdg4$<$\obs $<$23\fdg9 . Hence, as is reflected in Figure \ref{fig:obs}, without the HST data-point to serve as a reference position for the afterglow models in the fit, larger viewing angles cannot be ruled out (c.f. $18.3<$\obs$<20.3$ that is obtained when the HST data is considered). It may seem surprising that, of the two HST-omitted fits, the one without informative light-curve priors more precisely constrains the viewing angle, despite the fact that it results in a relatively flat posterior from \ratio$>11$ (similar to the Day 0, 75, 230 fit in Figure \ref{fig:geometries}). However, this is explained by the above-mentioned non-linear relationship between \ratio\ and \obs, where smaller angle ratios blow up the viewing angle; including the light-curve priors into the fit results in a \ratio\ posterior that peaks at slightly lower angle ratio values, resulting in slightly larger viewing angles.

In contrast to the day-0 data, we find that including the day-207 data pulls the viewing angle probability distribution toward larger values (by about $1\sigma$). This underscores the potential value that is added by each VLBI epoch, which mitigate the effects of using an imperfect jet model and imperfectly calibrated data (as will inevitably always be the case). That being said, not all VLBI observations post-merger (even hypothetically assuming identical quality and observing setup) hold equal value. For instance, we ran a fit with the day-230 data omitted, and found negligible difference in the resulting viewing angle measurement. Furthermore, as we have explained in Section \ref{res:scaling}, we have little ability to distinguish between jet energy distribution types because the first VLBI epoch occurs around 0.5\tp. For future events, obtaining at least one VLBI observation before 0.5\tp would enable significantly more discriminatory power between different jet angular power distributions. 

To summarize, an ideal GWM astrometry analysis would include at minimum 1) a sub-mas precision position of the kilonova, 2) a VLBI observation at roughly 0.25\tp, and 3) a VLBI observation close to \tp\ (but preferably several observations on and bracketing \tp). It may be impossible to meet the first criterion for GWMs that are unaccompanied by detectable kilonovae (due to e.g. extinction or a BH-NS that produces a jet but ejects  little neutron-rich matter). Such events would then benefit from multiple VLBI observations, especially at early times post-merger where the proper motion is greatest, although the faint radio emission during the early rising jet phase will require concerted observing effort. For nearly on-axis events, larger apparent superluminal motion is expected, but probing it would require dense VLBI sampling at early times post-merger. Identifying such events promptly would be a challenge and relies on characterization of the afterglow light-curve. Missing the light-curve peak would significantly weaken subsequent centroid constraints. Hence, a precise early-time reference position (e.g. from the kilonova or  VLBI observation as soon as the source is detectable) is essential to provide the astrometric anchor from which later centroid shifts are measured. \\

\subsection{Constraint on the viewing angle}
\label{disc:view}
Before including cosmological parameters, our most informed fit explores the full plausible jet geometry parameter space, includes all 4 astrometric datasets, and priors on \ratio\ and \tp\ that are informed by the well-studied afterglow light-curve (Section \ref{res:scaling}). It is the result from this fit (68\% credible interval: 18\fdg3$<$\obs$<$20\fdg3; 90\% credible interval: 17\fdg8$<$\obs$<$22\fdg0) that can be most directly compared to the constraint found by M22 (90\% credible interval of 19\degree$<$\obs$<$25\degree), because the same source distance has been assumed.  Our measurement is largely consistent with M22, but is 1.5 times more precise. \citet{govreen-segal23}, which first presented the model that is used in our analysis, fit their model to the image-plane flux-centroid displacements between the three VLBI epochs and the HST merger position from \citet{Mooley22}. They report \obs\ $=19\fdg4\pm2\fdg1$, which is notably consistent with our result, suggesting that 
more robust constraints are attained from full afterglow image modeling compared to approaches that use a point-source model approximation.

%the improved afterglow model relative to earlier studies (as outlined in Section \ref{meth:scaling}) is important.

These constraints, however, are all obtained by assuming the surface brightness fluctuation (SBF) distance of $40.7\pm2.4$\,Mpc from \citet{Cantiello18}, which was calibrated using the Cepheid distance scale. Hence, these constraints on viewing angle are inherently tied to the local distance ladder. We can obtain a viewing angle constraint that is independent of distance ladders by incorporating the gravitational wave information into our fit as described in Section \ref{meth:h0} and fitting for distance and peculiar velocity corrections directly. We find that including cosmology and GW information into the fit results in a slightly smaller viewing angle, 16\fdg8$<$\obs$<$19\fdg2, and $2\fdg1<$\cp$<4\fdg3$. Interestingly, the choice of galaxy group catalog and peculiar velocity reconstruction has little effect on the resulting posterior probability distribution of \obs. Indeed, all fits result in a $D_L$ posterior that peaks at 44\,Mpc, which in turn dictates the peak of the posterior of $H_0$, given a peculiar velocity correction. If we instead invoke an $H_0$ prior distribution informed by the SH0ES measurement, we find that larger viewing angles are preferred, which is more consistent with the M22 constraint. This is perhaps unsurprising given that the SH0ES measurement is of course calibrated using Cepheids (as is the case for the SBF distance invoked in M22). These viewing angle constraints are all summarized in Figure \ref{fig:obs}. 

We report a Bayesian model averaged luminosity distance of $44.0\pm1.6$\,Mpc (again, there is negligible resulting difference between peculiar velocity correction choice). This is more than $1\sigma$ larger than the SBF distance of $40.7\pm2.4$\,Mpc from \citet{Cantiello18}, when considering both uncertainties in quadrature sum. It is thus  important to carefully consider the luminosity distance to GW170817 when reporting viewing angle constraints.

%\textbf{Udi/Taya: Can we use our results to say something new/interesting about total grb energy and the Lorentz factor?} \textbf{TGS: I don't think so, let me talk to Udi about it.}
%As illustrated in Figure \ref{fig:obs}, this provides the most accurate and precise measurement yet on the viewing angle of GW170817.
%\textbf{Discuss the M22 analysys and source of this difference. Ultimtely should come down to having explored the full jet geometry parameter space and directly fitting the visibilities as opposed to only using the image-plane positions/displacements}.

\subsection{Hubble Constant measurement}
\label{disc:h0}
To assist with the discussion, we briefly summarize the significance of velocity corrections in standard siren measurements of $H_0$. We refer the reader to \citet{HowlettDavis20} and \citet{Nicolaou20} for a more in-depth discussion on this topic. $H_0$ describes the rate at which the Universe is currently expanding and the rate at which objects are receding from one another. In practice, an accurate measurement requires one to separate out all unrelated velocities from the object's total velocity. In the nearby Universe, such as at the distance to GW170817, the  `peculiar velocities' caused by the random motions of galaxies are a dominant source of systematics. The net peculiar velocity of a given galaxy is influenced by both the galaxy's (effectively random) motion within its own local galaxy group and the coherent motion of that group with respect to other nearby large-scale structures. Hence, it is standard practice to reduce the first of these by measuring/modelling instead the velocity of the \textit{group}, rather that those of the object in question. This approach then requires a choice of which galaxies should be included in the group and how to calculate the group's peculiar velocity. In their GW170817 standard siren  measurements of $H_0$, \citet{170817LVKh0} report  a result ($70.0^{+12.0}_{-8.0}$ \kms)\footnote{Where we quote $H_0$ values when EM information is not included, the maximum a posteriori (MAP) value and 68\% credible interval is reported. Where EM information is included, we report the median and 68\% credible intervals. The latter instances should have similar MAP values.} using one such combination of choices. \citet{Mukherjee21} also use a single combination but take a more sophisticated approach to deriving the velocity field that extends beyond linear perturbation theory to obtain $68.3^{+12.0}_{-8.0}$ \kms. \citet{HowlettDavis20} rigorously explore these systematics and assess the impact of using different plausible combinations of the aforementioned choices. In particular, they explore 154 different combinations and use Bayesian Model Averaging to report a measurement of $66.8^{+13.4}_{-9.2}$\,\kms. Similarly motivated, \citet{Nicolaou20} approach the peculiar velocity systematic by introducing Gaussian smoothing of galaxies in the 6dF Galaxy Survey \citep{springob14} as a nuisance parameter to determine a single peculiar velocity correction and obtain $H_0=68.6^{+14.0}_{-8.5}$\,kms. 

In addition to peculiar velocities, the degeneracy between GW170817's inclination angle and distance is a dominant source of uncertainty on these standard siren measurements. As described in the introduction section, several studies have attempted to break this degeneracy by invoking centroid motion information of the afterglow. Generally, for a given peculiar velocity correction, a systematically larger inferred viewing angle will result in a systematically smaller inferred luminosity distance and hence a larger inferred $H_0$ value. Here, we provide a comparison of a representative sample of key $H_0$ results from the literature and their associated jet viewing angle constraint. \citet{hotokezaka19} fit for 5 to 7 jet model parameters (for two different jet-types), including \obs\ and $D_L$, using the image-plane centroid displacement between days 75 and 230 from M18, as well as the afterglow light-curve. The models however are purely synthetic and derived semi-analytically, to allow parameter space to be easily explored through MCMC at the expense of correct modeling of the fluid dynamics. The posterior probability distributions of \obs\ and $D_L$ obtained using strictly the GW data by \citet{170817LVKh0} were then used as their priors, resulting in \obs\ posteriors peaking at approximately $17\degree$ for both jet-types considered. They then follow the exact same $H_0$ inference and peculiar velocity methodology as \citet{170817LVKh0} to measure $H_0=68.9^{+4.7}_{-4.6}$\,\kms. M22 present a revised $H_0$ measurement of $71.5^{+4.6}_{-4.6}$\,\kms, where they have followed an identical peculiar velocity/$H_0$ strategy to \citet{hotokezaka19} and \citet{170817LVKh0}, but have added the day-207 and HST data into their fit, yielding a systematically larger viewing angle posterior ($90\%$ credible interval of 19\degree to 25\degree). To include EM information in their own analysis, \citet{HowlettDavis20} take the \obs\ and $D_L$ posteriors from \citet{hotokezaka19} (i.e. using only days 75 and 230 astrometric data) to obtain $64.8^{+7.3}_{-7.2}$\,\kms\ using their Bayesian Model Averaging approach. \citet{Mukherjee21} combine the same dual-epoch HSA VLBI information with their obtained peculiar velocity correction ($373\pm130$\,km\,s$^{-1}$) by using a flat prior on inclination angle of 0.25 rad $<\imath\frac{D_L}{41\,\text{Mpc}}<0.45$\,rad to produce $H_0=68.3^{+4.6}_{-4.5}$\,\kms. For all other measurements quoted above that use reconstructions, a peculiar velocity correction uncertainty of $150$\,\,km\,s$^{-1}$ was assumed. \citet{HowlettDavis20} marginalize over a variety of peculiar velocity catalogs where the peculiar velocity measurements in and around the group are averaged over, for a given catalog. The corresponding uncertainties thus depend on the particular combination of galaxy group and peculiar velocity catalog, and range from approximately $100-750$\,km\,s$^{-1}$ (see their Table 2).  

In our own analysis, the $H_0$ measurement strategy is drawn from \citet{HowlettDavis20}, though we have used new and improved peculiar velocity reconstruction maps and assume an uncertainty of 150\,km\,s$^{-1}$ (see Section \ref{meth:h0}). We have considered 28 combinations of galaxy group catalogs and peculiar velocity corrections. We find that these choices can systematically adjust the $H_0$ measurement by up to roughly $1\sigma$ ($\sim4$\,\kms), and that there is little difference between their Bayes evidences (the largest Bayes factor in natural logarithm space is 1.8). These points emphasize that the peculiar velocity correction remains an important systematic in GW-VLBI $H_0$ constraints, particularly for the most nearby events. We also investigated the sensitivity to the somewhat arbitrary albeit standard 150\,km\,s$^{-1}$ peculiar velocity uncertainty by using a more conservative value of 250\,km\,s$^{-1}$ and found the resulting peak or median $H_0$ posterior values to decrease by less than 1\%. As illustrated in Figure \ref{fig:H0}, our Bayesian model averaged $H_0$ value across these peculiar velocity corrections is more consistent with the \citet{planck20} $H_0$ value than that measured by \citet[][SH0ES collaboration]{reiss22}. Specifically, our $H_0$ value is within $0.5\sigma$ of the Planck result and is $1.7\sigma$ lower than the SH0ES measurement.
%caution of the potential presence of model mispecification

It is tantalizing that our independent late-Universe $H_0$ measurement is more consistent with Planck's early-Universe measurement. Taken at face-value, this result could challenge the potential for non-standard cosmology models to resolve the Hubble tension. We now discuss possible caveats to our measurement. First, we consider the potential presence of model mispecification. In particular, while most existing jet models are able to consistently and correctly predict the afterglow close to \tp, the late-time evolution ($\gtrsim 350$ days post-merger) of the light-curve is not well fit by standard synchrotron afterglow models (including those used in this study; e.g., \citealt{Gianfagna24}). The source of observed excess emission above what is predicted at these late times is currently unknown. In particular, it is unclear whether information is missing from the jet models themselves or the flux excess stems from an unrelated energy source \citep[such as the merger remnant or a kilonova synchrotron afterglow;][]{Troja22,Hajela22,Dastidar24,Dupont24,Katira25}. \citet{Gianfagna24} find that for standard jet models, larger viewing angles are required to force consistency with the late-time light-curve, which results in unrealistically large $H_0$ values. Such model mispecification concerns may be alleviated by the fact that, at the times post-merger probed by the data used in our analysis, it is the geometry that dominates the evolution of the observed afterglow over any model inaccuracies that may be present. More generally, that we find a similar result (at the $\sim1\%$ level) with uninformed priors on \ratio\ and a range of values for \tp\ (see Section \ref{res:h0}) implies that our strategy is robust against some of the complex systematics associated with modeling the afterglow light-curve, at least within the times post merger probed in our analysis. Still, we caution that the level of fidelity of the jet models that we have used in our analysis could impact our results. The jet afterglow of GW170817 remains the best studied example for any SGRB/BNS; the presence and extent of any significant model mispecification in this analysis is impossible to ascertain without additional well-studied afterglows that can serve to test and refine existing models. Second, we note that all of the analyses discussed in this subsection (including our own) have made the standard assumption that the relativistic jet is launched along the angular momentum axis and that this is perpendicular to the orbital plane. \citet{Muller24} considered the consequences should this assumption be wrong, where greater misalignments ($i-$\obs) would result in smaller $H_0$ values in a non-linear fashion.  Given the assessment of this non-linear bias that is presented in \citet{Muller24},  we estimate that a misalignment of $\sim-10$\degree\ would be required to systematically push the peak of our $H_0$ posterior to the larger SH0ES value.
 %It is unclear whether (and to what extent) this lack of understanding impacts the jet models used in this analysis.
 
 %To assess the impact of the complex systematics associated with the light-curve on our measurement, we run three additional $H_0$ fits with flat priors on \ratio\ and \tp\ set to 145, 155 and 165 days and obtain  

 %We investigate the sensitivity of our results to \tp\ by repeating the fit with fixed luminosity distance and uninformative \ratio\ and \diff\ priors that is described in Section \ref{res:scaling}, but changing \tp\ from 155 days to 145 and 165 days. We find similar maximum a posteriori values for the jet geometry across all three \tp\ choices, which further indicates the robustness of our results to uncertainties related to modeling of the light-curve.

%More well-sampled binary neutron star merger afterglows are required to address these questions.

At this point, it is instructive to  assess the relative contributions of the various sources of uncertainty on our $H_0$ measurement. The fractional uncertainty contributed by the peculiar velocity is 0.05. The GW uncertainty contribution due to the GW signal's detection significance ($32.4\sigma$) is 0.03. Finally the contribution due to the uncertainty of 1\degree\ around a viewing angle of 18\fdg0 is about 0.01. The total uncertainty is $\sim6$\% (or 4.4\,\kms) of $H_0$ and we therefore can see that the dominant source of uncertainty for inferences based on GW170817 is the peculiar velocity of its host galaxy. It is for this reason that all studies with a reasonable constraint on the viewing angle report similar uncertainties, while the differences in the mean $H_0$ values stem primarily from the different peculiar velocity models. Future, more distant events will suffer proportionately less from peculiar velocity uncertainties, but this will come at the cost of reduced precision on the viewing angle measurement.

\section{Conclusion and outlook}
\label{Conclusion}
%I need to explain somewhere that our method allows us to marginalize over the astrometric systematic uncertainties (not clear to me how this was addressed in M22)

The detection of GWs from binary neutron star mergers provides an independent way of measuring $H_0$. Before considering the electromagnetic counterparts caused by the relativistic jet, the inherent covariance between the inclination of the orbital system and its luminosity distance was the dominant source of uncertainty in the first and only bright siren measurement of $H_0$ using GW170817. We have presented a rigorous Bayesian framework that enables a continuum of jet model geometries to be fit directly to VLBI visibilities of the jet afterglow, to directly constrain the inclination of the GWM. We have used this framework to fit the full multi-epoch milliarcsecond-scale astrometric dataset that exists for GW170817. When we include information about the afterglow light-curve, we obtain the most informed measurements of GW170817's viewing angle.  Specifically, we find $18.3<$\obs$<$20\fdg3 (for $D_L=40.7$\, Mpc), which is largely consistent with the measurement reported in \citet{Mooley22} but more precise, and consistent with the constraint reported by \citet{govreen-segal23} where the same model as this study was used. We have expanded the framework by including relevant velocity and GW information to directly measure $H_0$. To address the dominant $H_0$ measurement systematic induced by uncertainties in the peculiar velocity correction, we have considered various combinations of galaxy group catalogs and peculiar velocity reconstruction maps. We have used Bayesian Model Averaging to consider these plausible corrections and have obtained $H_0=65.5\pm4.4$\,\kms. Notably, this value is in better agreement with the early-universe Planck value (within $0.5\sigma$) than the late-Universe SH0ES value, where there is a $1.7\sigma$ discrepancy. The consistency of our independent late-Universe measurement with the early-Universe method may pose a challenge for time varying cosmology models. We have discussed possible caveats associated with our result and, ultimately, more standard siren measurements will be required to increase the significance of this finding.

%We have presented a rigorous Bayesian visibility-plane modelfitting framework that, for a given jet model, allows for the full plausible range of relativistic jet geometries to be compared to VLBI data of jet afterglows. We have expanded this framework to directly measure $H_0$, by including relevant velocity and GW information.

%We report $H_0=61.8^{+4.1}_{-3.9}$\,\kms and $H_0=65.7^{+4.2}_{-4.0}$\,\kms, depending on the choice of galaxy group catalog and peculiar velocity reconstruction method, and emphasize that peculiar velocities remain an important systematic in bright siren analyses.  Nevertheless, our $H_0$ results are  much more consistent with the \citet{planck20} measurement than that of the SH0ES collaboration \citep{reiss22}. 

Our results have highlighted the importance of considering the full jet geometry parameter space in such analyses and the constraining power that is gained when directly fitting the VLBI visibilities (which is more accurate and adds information about the source-size to the fit). We have assessed the relative value added by each of the four astrometric datasets available to help guide the VLBI follow-up strategy of future EM-bright off-axis GWM events. We emphasize that our model-fitting framework is generic and can be applied to future GW events as well as GRBs more generally, but may require different hydrodynamical models for on-axis events.

We have demonstrated the power that VLBI measurements of the jet afterglow possess to reduce the GWM inclination angle--luminosity distance degeneracy. In GR, the inclination angle can be obtained directly if the two GW polarizations are precisely measured, though there could be up to six polarization modes in alternate gravity theories. This would be preferable to relying on the VLBI method, as the analysis would be independent of EM data systematics and imperfect jet models. In practice, however, it is unlikely that polarization measurements will achieve better precision compared to the VLBI method in the near future \citep{Burns2020}. This is because a GWM event has to be observed by $n$ detectors to resolve $n$ polarization modes and, even for $n=2$ events and assuming only the two GR polarization modes, there is substantial correlation between the modes' amplitude parameters \citep[e.g.][]{Takeda18}. Currently, the two most sensitive detectors are co-aligned, which precludes their independence in this context. A detection by all of the 5 second-generation ground based detectors (LIGO-Hanford, LIGO-Livingston, Virgo, KAGRA, and LIGO-India) could provide helpful inclination constraints. However, the realization of such a scenario will be limited by the least sensitive detector in the network as well as the combined duty-cycle of the detectors. Compounding this, the two polarizations are more difficult to distinguish for face-on events (i.e. the loudest and hence most detectable GWM events) given their cos$\imath$ terms. Therefore, high spatial resolution observations of the relativistic outflows of GWMs are expected to remain uniquely powerful, albeit rare, tools for high-precision bright siren $H_0$ measurements until at least the next generation of ground-based detectors (3G) become operational, and would continue to provide valuable independent information thereafter.

We estimate that about a dozen bright siren measurements similar to the one from this analysis would be required to achieve an $H_0$ precision of 2\%, assuming that each bright siren measurement is independent. However, peculiar velocities are correlated and a complex statistical solution is required to solve the covariance matrix in order to properly combine multiple standard siren measurements \citep{BlakeTurner24}. At distances below 100\,Mpc, this may double the true number of sirens that are required to reach the aforementioned precision.  It is critical, then, that the extractable information from each rare future GW-VLBI dataset is maximized, such as we have set out to accomplish in this study.
%for breaking the $cosi$--$D_L$ degeneracy.
 
%\section{Software and third party data repository citations} \label{sec:cite}

\begin{acknowledgments}
ATD, KG, KM and EN acknowledge support through
DP200102243. CH acknowledges support through ARC Dis-
covery Project DP20220101395. Parts of this research were conducted by the Australian Research Council Centre of Excellence for Gravitational Wave Discovery (OzGrav), through project number CE230100016. EN and TGS acknowledge support by an ERC consolidator grant JetNS (818899). We thank Eric Thrane for insightful discussions and advice on Bayesian statistical methods. 
\end{acknowledgments}

\software{\textsc{Bilby} \citep{Bilby},
          \textsc{GALARIO} \citep{galario},
          \textsc{difmap} \citep{difmap},
          \textsc{AIPS} \citep{aips},
          \textsc{parseltongue} \citep{parseltongue},
          \textsc{Dynesty} \citep{Dynesty},
          %\textsc{corner} \citep{corner}
          \textsc{GetDist}
          \citep{getdist}}

\appendix

\section{Re-analysis of Mooley et al. 2018}
\label{M18}
\subsection{Fits to HSA VLBI data only}
\label{res:free}
To validate and assess the usefulness of the model fitting framework above, we consider the set of five model afterglow images that were originally presented and analyzed in M18. These correspond to four models from simulation A in M18, a power-law jet with core opening angle at peak-flux of 0.08 radians and with viewing angles of 0.25, 0.35, 0.45 and 0.50 radians; and one model from simulation B, which has an opening angle at peak-flux of 0.06 radians and viewing angle of 0.30 radians (see extended data Table 2 of M18 for further details about these models). Each of these five models has a pair of afterglow model images, corresponding to day 80 and day 240 post-merger (which are not actually exact matches to the average times of the two VLBI HSA datasets, but were deemed sufficiently consistent in M18). Each model image consists of $100\times100$ pixels of side length 0.2\,mas. 
%For the prior range on the astrometric uncertainty nuisance parameters (\delra\ and \deldec) we use the astrometric uncertainties reported in M18 on the compact check source, NGC 4993, located in the field. Across 7 HSA observations (our first HSA epoch consists of the concatenation of 3 observations, and the second consists of the concatenation of 4 observations), the root-mean-square variation on the position of the source is 0.14\,mas in RA and 0.49\,mas in Dec. These values are used to construct Gaussian priors for \delra\ and \deldec.  We assume a ten per cent fractional error on the flux scale of each VLBI dataset (as estimated in  M18). Uninformative flat priors are used for all other parameters. Detailed results from this re-analysis are presented in Section \ref{res:free} of the Appendix.
%We consider the five afterglow models that were originally fit to the days 75 and 230 post-merger HSA VLBI data in M18, as described in Section \ref{meth:fit1}. 
The resulting log-evidences are presented in Table \ref{tab:fits} of the Appendix. The astrometry plots presented in Figure \ref{fig:hstfit} compare the afterglow models, after applying the median posterior values of the parameters described in Section \ref{methods}, to the HSA image-plane centroid positions of GW170817, to help visualize goodness of fit. The latter values were derived using the \textsc{AIPS} \textsc{jmfit} task, which fits a single 2-D Gaussian component to interferometric data in the image-plane. The images themselves were produced by fitting a single circular Gaussian model component to the visibilities. We have not used those \textsc{jmfit} values reported in M18 and M22, because these were affected by a now-corrected rounding bug in  \textsc{difmap} that had underestimated the beam size by 6.6\%.\footnote{\textsc{difmap} change-log, 05/20/2019 entry. \url{ftp://ftp.astro.caltech.edu/pub/difmap/change_details.txt}} The correct values are reported in Table \ref{tab:astrometry}.

Model A35\footnote{The number in these model names corresponds to the viewing angle in radian units.} (\obs$=20$\degree) has the largest log-evidence and is used as the reference model to calculate the Bayes factors ($B$) in log-scale that are reported in Table \ref{tab:fits}. We provide a qualitative interpretation of $B$ between models, following \citet{Jeffreys61}. Of all models, A50 (\obs$=29$\degree) is the least favored, with a Bayes factor (in natural logarithm space) of 2.63, as the motion predicted between the VLBI epochs is clearly too low, as depicted by the yellow data-points in Figure \ref{fig:hstfit}. There is moderate evidence against A25, which predicts too much proper motion between the VLBI epochs. As a result, the position angle posterior of this model is relatively large, to counteract this fact and attempt to align the model flux centroids with the data along the right ascension axis (where the HSA angular resolution and hence positional precision is highest). Besides these latter two models, the remaining models (A35, A45 and B30) fit the data statistically just as well.

To compare these results to those obtained in M18, we briefly summarize the model comparison strategy of M18. The five 2-D afterglow models were fit to the data by representing their non-zero pixels as a collection of delta function model components in the image-plane and fitting these to the observed visibilities by brute-force using the Levenberg-Marquardt non-linear least squares minimization technique with \texttt{modelfit} in \textsc{difmap}. There, RA, Dec, PA and flux scale were taken as free parameters.  To assess the goodness of their fits, the resulting minimum $\chi^2$ value found for each model was compared to the $\chi^2$ values obtained when fitting a singular circular Gaussian model component and perturbing the resulting fitted model in RA and Dec by 0$\sigma$, 1$\sigma$, 2$\sigma$ and 3$\sigma$ in both directions and recomputing $\chi^2$ values for each permutation. This enabled the $\chi^2$ values of the afterglow models to be assessed for consistency with the best-fit circular Gaussian model. In each $\chi^2$ calculation instance, the observed visibilities were allowed to vary in RA and Dec by their systematic uncertainties. Within this framework, model A35 resulted in a $\chi^2$ value comparable to a best-fitting circular Gaussian perturbed by roughly $1\sigma$ and B30 was consistent within $2\sigma$, whereas the other 3 models had $\chi^2$ values beyond a corresponding $2\sigma$ perturbation. Thus, model A35 was claimed to best fit the data with B30 not possible to rule out.

Therefore, the findings from M18 are generally consistent with our more refined approach. Our results do, however, show that A45 is not statistically significantly disfavored over A35 and B30. Furthermore, we find that A25 was more strongly disfavored in M18 than is actually suggested by our results. The posteriors for this model reveal a relatively strong covariance between the PA of the model and both the reference position RA and Dec offsets. Upon investigation of the M18 analysis of A25, we determined that the range of model position angles explored was not sufficiently large, such that the true $\chi^2$ minimum was never actually found in M18's brute force parameter exploration.

\subsection{Inclusion of Hubble Space Telescope kilonova astrometry}
\label{res:hst}

%\citet{Mooley22} (M22) perform precise astrometric measurements of GW170817 using Hubble Space Telescope (HST) observations of its optical counterpart 8 days post-merger, when the emission was dominated by the kilonova, to effectively characterize the position of the neutron star merger. M22 transform the HST data into the GAIA/ICRF3 frame and similarly derive offsets for the VLBI data, which is in the frame of phase calibrator source J1312-2350, relative to the ICRF3 frame. Thus,

\begin{figure}[h]
\plotone{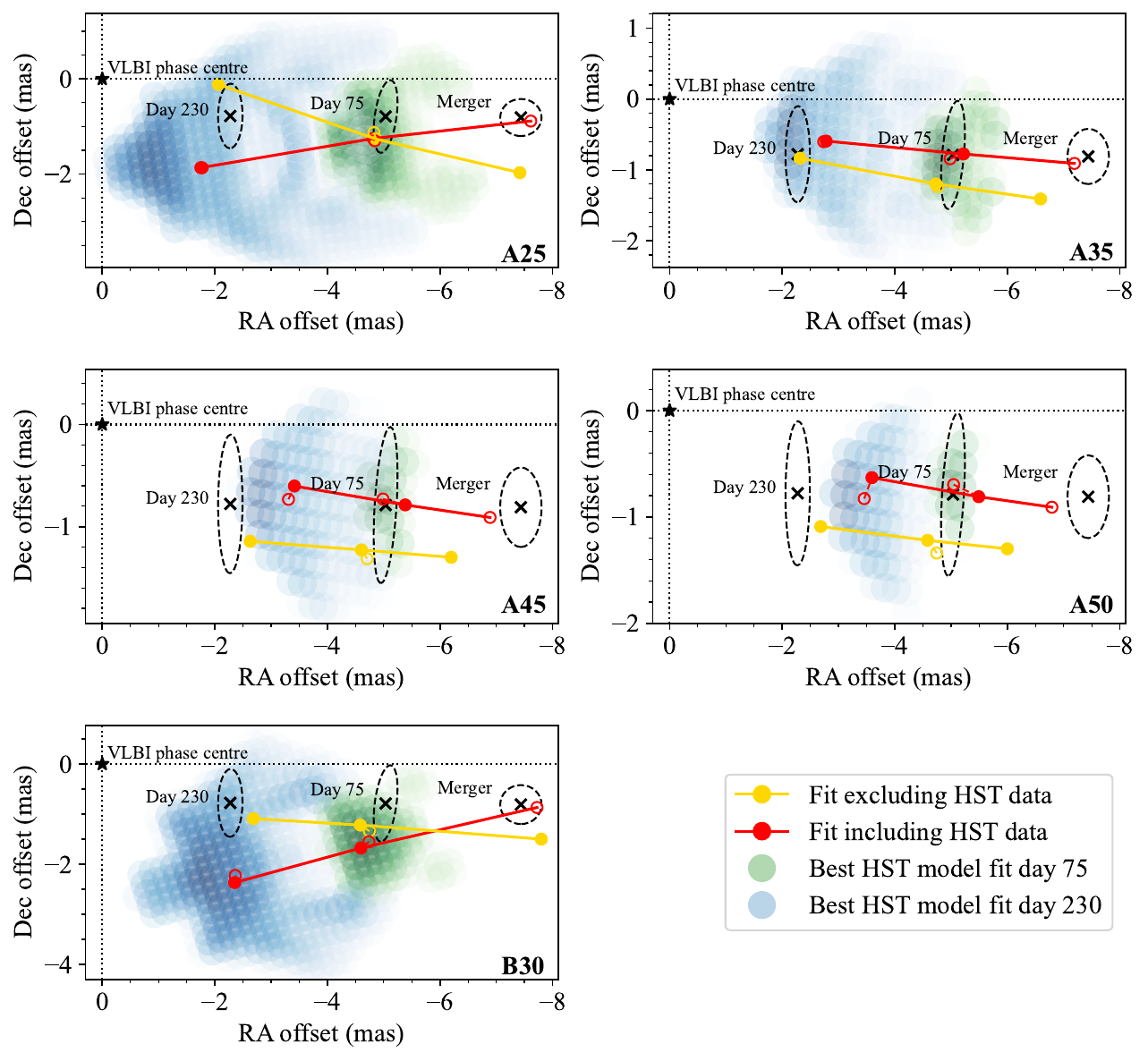}
\caption{Reanalysis of the afterglow models originally analyzed in \citet{Mooley18} (labeled in the bottom-right corner of each panel). The black X-shaped markers are the image-plane positions of GW170817 with their uncertainty ellipses and are the same as described in the caption of Figure \ref{fig:scaling}. The closed yellow points correspond to the resulting best fit position (the flux centroid of the fitted model) when the model reference position offset (\ra, \dec) is a free parameter with uninformative priors. The open yellow circle denotes the posterior probability of the astrometric nuisance parameter (\delra,\deldec) pair associated with the VLBI data (with priors that capture the astrometric systematic uncertainties of both epochs added in quadrature, as opposed to fitting two separate pairs). The red data points correspond to the same but when the fit is anchored to the HST kilonova/merger position (Section \ref{res:hst}). The closed red VLBI circles represent the flux centroid of the model referenced to the fitted HST position (right-most open red circle; (\ra, \dec)). The open red VLBI circles include the respective posterior probabilities of the VLBI astrometric nuisance parameters, and thus represent where the centroid would appear on the sky after including the systematic error inferred for the data on that date. In each panel, the corresponding afterglow model-images are represented by shaded green (Day 75) and blue (Day 230) background circles. The shaded models shown in the background are plotted using the latter best fit positions.} %We note that, counter-intuitively, the HST reference position (right-most solid red circle; before including posteriors of the HST astrometric uncertainty captured by \ra\ and \dec) does not overlap exactly with the black HST data point. This is caused by the finite pixel size (0.2\,mas) of the merger location in the models, which, when rotated by the best-fitting PA, results in an apparent offset of 0.14\,mas.} 
\label{fig:hstfit}
\end{figure}

As described in Section \ref{meth:hst}, M22 provide astrometric information that enables a joint analysis of the HST and HSA VLBI data, where the HST data provides a measurement of the location of the merger itself, thus specifying the reference position offset for the M18 models. The five dual-epoch pairs of M18 models are fit to the VLBI data according to Section \ref{meth:hst}, with the reference position offset anchored to the HST merger position measurement. The resulting astrometry plots for the five M18 models are presented in Figure \ref{fig:hstfit} (red data-points). There, it is evident that none of the models are good fits to the joint HST-HSA dataset. In particular,  whereas A35 (the best-fitting model reported in M18) fit the data reasonably well in the previous subsection, now there is clearly not enough angular displacement between the two VLBI epochs, once the the HST measurement which anchors the day-0 position is included. The resulting log-evidences (Table \ref{tab:fits}) of the five model-fits indicate that models A25 and B30 are now strongly preferred, likely because those models contain the largest angular displacement between the two epochs, whereas the other 3 models have insufficient motion. However, as is evident in the astrometry plots for A25 and B30, the fitted PAs change significantly, as the fit tries to force consistency with the observed motion in right ascension, at the expense of the goodness-of-fit in the less-precisely constrained declination direction. Thus, it appears that these five models do not adequately sample the full plausible range of jet parameter space. We consider this to be further motivation for the analyses that are performed in Section \ref{results}.

%An additional pair of astrometric nuisance parameters is introduced to account for the uncertainty in the position of the VLBI reference source (used to bring the HST position from the GAIA/ICRF3 frame into the HSA VLBI frame). The likelihood is also adapted to include an additional pair of chi squared terms to account for the statistical uncertainty on the RA and Dec of GW170817 in the HST data.

%The resulting best fit for F35 is show in Figure \ref{fig:hstfit}. It is clear, here, that when anchoring the model reference position to the HST merger location measurement, F35 is not a good fit to the joint HST-VLBI dataset; there is not enough motion between the first epoch and the time of merger. We fit new sets of models, exploring a much wider set of parameter space in the next subsection.
%The open yellow circles include the posterior probabilities of the astrometric nuisance parameters that correspond to the respective systematic uncertainties in RA and Dec

\begin{figure}[h]
\centering
%\plotone{GW_KDE_corner_pub.png}
\includegraphics[width=1.0\textwidth]{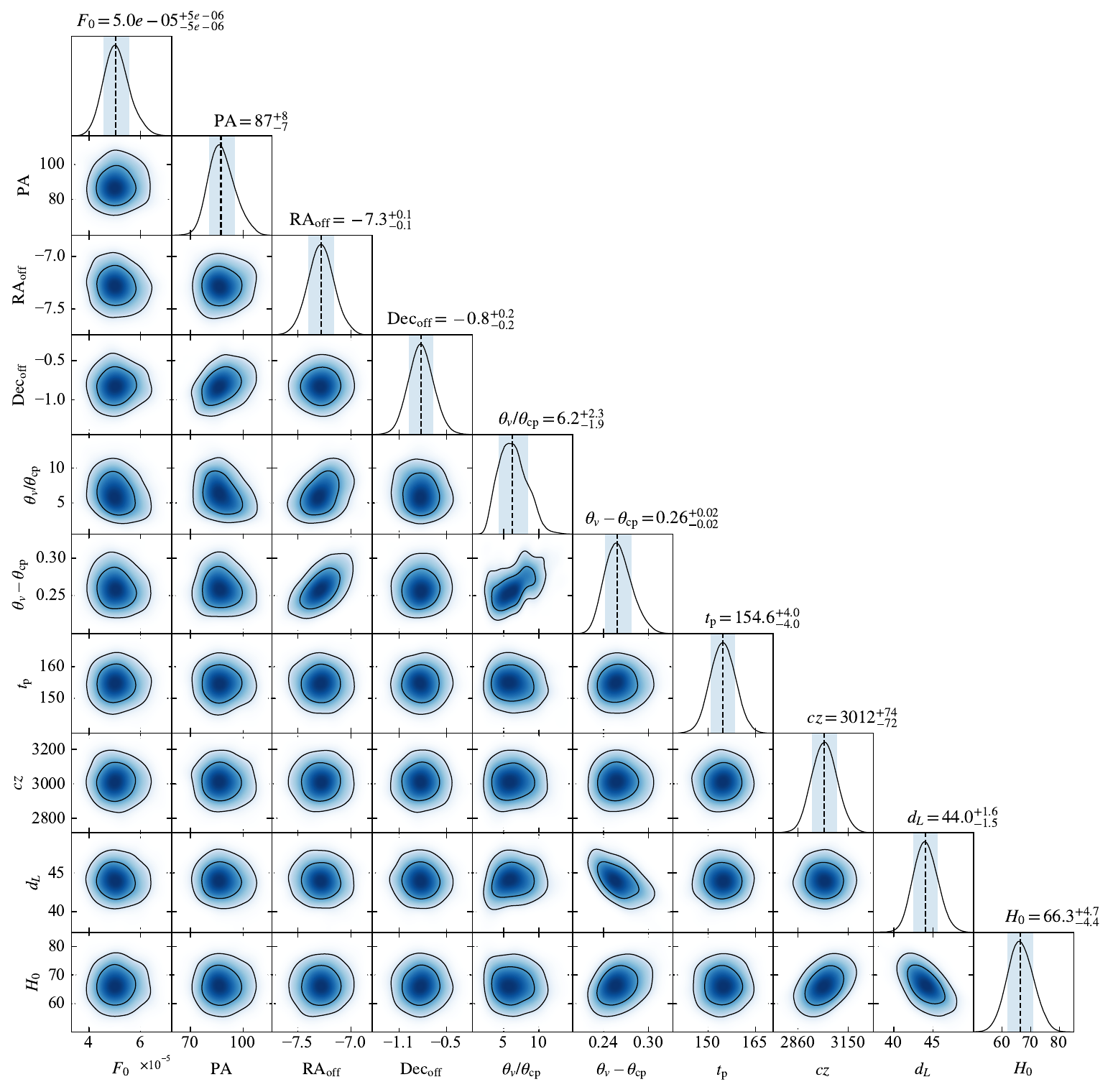}
\caption{Example corner plot of key model parameters' posteriors of a fit from Section \ref{res:h0}. The parameters are described in Table \ref{tab:params} and Section \ref{methods}. These results are generated using the peculiar velocity correction model with the largest Bayes evidence of the 28 corrections that were considered in Section \ref{res:h0}. The posterior median and 68\% credible interval are reported above each histogram and are denoted in the histograms by a vertical dotted line and shading, respectively. The corner plot contours correspond to the 68\% and 95\% credible intervals. } 
\label{fig:corner}
\end{figure}

\begin{table}[]
\centering
\caption{Log-evidences (ln($z$)) and Bayes factors ($B$, in natural logarithm space) of various fits performed in this analysis. The ``Free fit'' column corresponds to the fits presented in Section \ref{res:free}, where the reference position offset of the M18 models is a free parameter. The ``HST fit'' column contains results corresponding to Section \ref{res:hst}, where the reference position offset of the M18 models is locked to the HST kilonova position measurement. The ``Model scaling'' table corresponds to the HSA-HST (days 0, 75, 230) fits from Section \ref{res:scaling}, where the full plausible parameter space of jet geometry is explored for a given jet type, in each fit (PL stands for `power law'). In the left-hand table, the number in each of the model names corresponds to the model's viewing angle, in radians. We follow \citet{Jeffreys61} in our interpretation of the Bayes factors: $B\gtrsim5$ corresponds to decisive evidence against the null hypothesis, $2.3\lesssim B \lesssim5$ denotes strong evidence, $1.1\lesssim B\lesssim2.3$ signifies substantial evidence, and $B\lesssim1.1$ denotes insignificant evidence against the null hypothesis.}
\label{tab:fits}
\begin{tabular}{ccc|cc}
                         & \multicolumn{2}{c|}{\textbf{Free fit}} & \multicolumn{2}{c}{\textbf{HST fit}} \\ \hline
                         & ln($z$)              & $B^a$           & ln($z$)            & $B^b$             \\ \hline
\multicolumn{1}{c|}{A25} & $-47069.97(7)$           & 1.57            & $-47064.76(5)$                 & -           \\
\multicolumn{1}{c|}{A35} & $-47068.40(7)$           & -               & $-47067.40(6)$                 & 2.64        \\
\multicolumn{1}{c|}{A45} & $-47068.87(7)$           & 0.47            & $-47077.86(7)$                 & 13.1       \\
\multicolumn{1}{c|}{A50} & $-47071.03(7)$           & 2.63            & $-47085.27(8)$                 & 20.5          \\
\multicolumn{1}{c|}{B30} & $-47068.89(7)$           & 0.49            & $-47065.14(5)$                 & 0.38          \\
\hline
\end{tabular}
\quad
\begin{tabular}{lccc}
\multicolumn{4}{c}{\textbf{Model scaling fit}}       \\ \hline
\multicolumn{1}{c}{}            & ln($z$) & $B^c$ & \diff \\ \hline
\multicolumn{1}{l|}{PL ($b=3$)} & $-47069.47(6)$     & 1.38 & 15\degree$^{+1}_{-2}$   \\
\multicolumn{1}{l|}{PL ($b=4$)} & $-47069.12(6)$     & 1.03 &  15\degree$^{+1}_{-2}$ \\
\multicolumn{1}{l|}{PL ($b=6$)} & $-47068.48(6)$     & 0.39 & 15\degree$\pm1$ \\
\multicolumn{1}{l|}{Tophat}     & $-47068.09(6)$     & - & 15\degree$^{+2}_{-2}$  \\
\hline
\end{tabular}

\raggedright
\vspace{0.5em}
\hspace{2.5cm}$^a$ Bayes factor relative to A35.\\
\hspace{2.5cm}$^b$ Bayes factor relative to A25 model.\\
\hspace{2.5cm}$^c$ Bayes factor relative to tophat model.
\end{table}

\begin{comment}

\begin{tabular}{ccc|cc|lccc}
                         & \multicolumn{2}{c|}{\textbf{Free fit}} & \multicolumn{2}{c|}{\textbf{HST fit}} & \multicolumn{4}{||c}{\textbf{Model scaling fit}}       \\ \hline
                         & ln($z$)              & $B^a$           & ln($z$)            & $B^b$            & \multicolumn{1}{||c}{}            & ln($z$) & $B^c$ & \diff \\ \hline
\multicolumn{1}{c|}{A25} & -47069.778           & 1.6            & -47064.786                 & -               & \multicolumn{1}{||l|}{PL ($b=3$)} & -47069.586     & 1.2 & 15\degree$\pm1$   \\
\multicolumn{1}{c|}{A35} & -47068.230           & -               & -47067.086                 & 2.3                & \multicolumn{1}{||l|}{PL ($b=4$)} & -47069.212     & 0.85 &  15\degree$^{+1}_{-2}$ \\
\multicolumn{1}{c|}{A45} & -47068.727           & 0.50            & -47077.718                 & 12.9               & \multicolumn{1}{||l|}{PL ($b=6$)} & -47068.661     & 0.30 & 15\degree$\pm1$ \\
\multicolumn{1}{c|}{A50} & -47071.890           & 3.7            & -47086.321                 & 21.5               & \multicolumn{1}{||l|}{Gaussian}   & xxx    & xxx & xx \\
\multicolumn{1}{c|}{B30} & -47068.813           & 0.58            & -47065.092                 & 0.31               & \multicolumn{1}{||l|}{Tophat}     & -47068.361     & - & 15\degree$^{+1}_{-2}$  \\
\hline
\end{tabular}
\end{comment}

\section{Model-fitting figures and tables}
\label{extras}
\begin{figure}[h]
\centering
%\plotone{GW_KDE_corner_pub.png}
\includegraphics[width=0.6\textwidth]{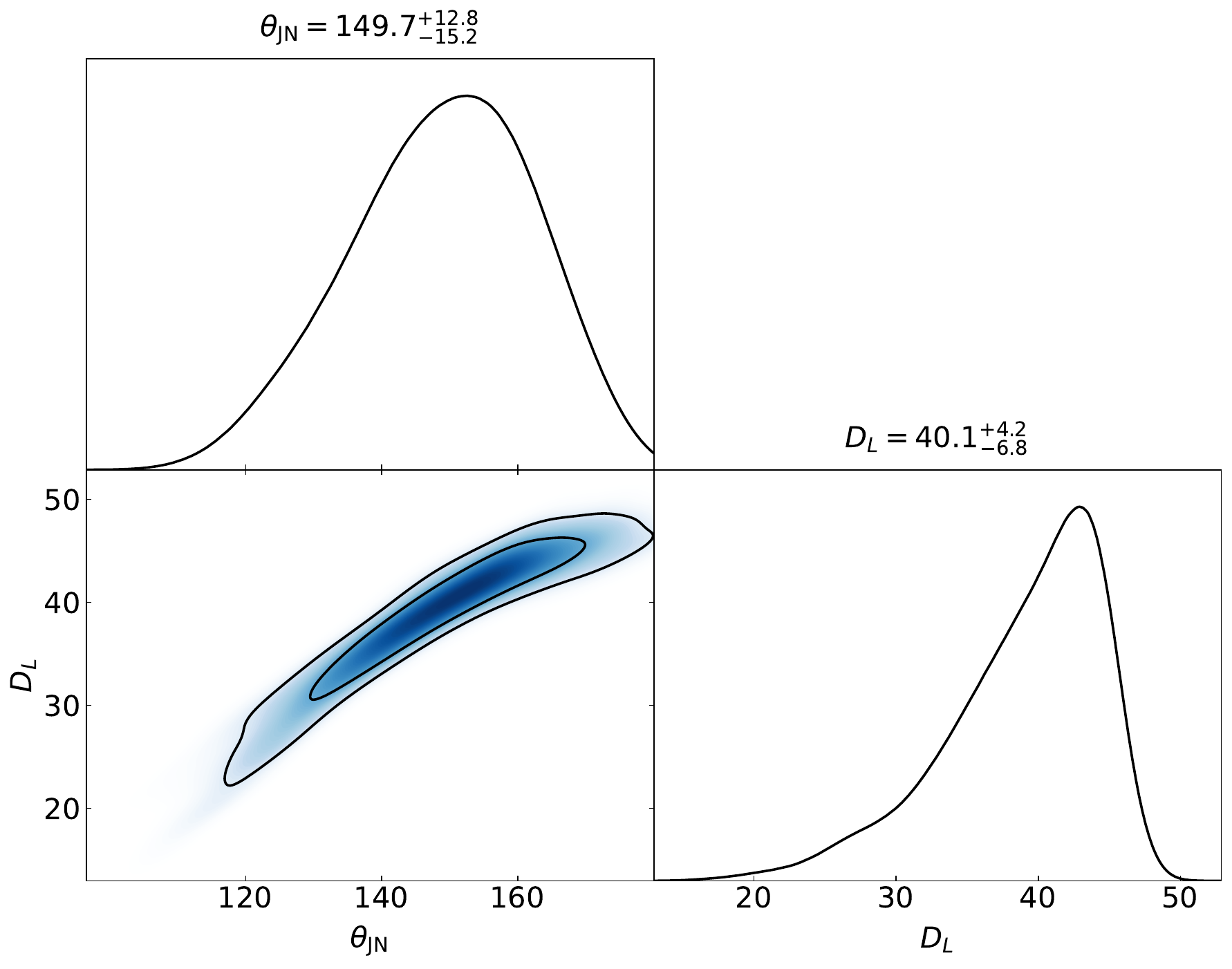}
\caption{Corner plot obtained by sampling from the GW inclination angle ($\theta_{\text{JN}}=\text{acos}(\imath)$, in degrees) and luminosity distance (in Mpc) KDE that was constructed from the posteriors reported in \citet{170817props}. This KDE is used for the GW likelihood described in Section \ref{meth:h0}.} 
\label{fig:kde}
\end{figure}

In this section of the Appendix, we include additional information related to the data used and the Bayesian fits preformed in this analysis. In Figure \ref{fig:kde}, we show the corner plot of the $\theta_{\text{JN}}$ and $D_L$ posteriors that result from evaluating the GW likelihood that samples the GW KDE explained in Section \ref{meth:h0}. This KDE is constructed, in the same way as \citet{HowlettDavis20}, from the \texttt{lowspin} posterior samples reported in \citet{170817props}.
Table \ref{tab:params} provides a summary of the parameters (and their priors) that are included in the various fits undertaken in this study.  We report the Bayesian model average (taken across all 28 $H_0$ peculiar velocity correction fits) posterior medians and 68\% credible intervals for all parameters in Table \ref{tab:posterior}. Figure \ref{fig:corner} provides an example corner plot from the most comprehensive fit of our analysis (Section \ref{meth:h0}). Astrometric uncertainty and flux uncertainty nuisance parameters have been omitted from the figure to improve legibility.
Lastly, we provide a summary of the centroid positions and their uncertainties for the astrometric data used in this analysis in Table \ref{tab:astrometry}.

\begin{table}[]
\caption{GW170817 positions and their uncertainties. All positions are reported in the frame of the VLBI phase reference calibrator source J1312$-$2350. All VLBI flux centroid position measurements were obtained by fitting a 2-D Gaussian in the image-plane (\textsc{AIPS} task \texttt{JMFIT}). We use astrometric information reported in \citet{Mooley22} for the HST position and its uncertainties. (1) The mean number of days post-merger that the position corresponds to; (2) The telescope used for the corresponding position measurement; (2) the coordinates of the position of GW170817 in the J1312$-$2350 frame (RA, Dec); (4) the statistical uncertainty on the measurement in mas (RA, Dec). For VLBI data this corresponds to the uncertainty of the 2-D Gaussian fit as reported by \texttt{JMFIT}; (5) The systematic uncertainty on the measurement in mas (RA, Dec). For VLBI data, this corresponds to the uncertainty caused by the ionosphere. For HST, this corresponds to the uncertainty in the position of J1312$-$2350, used to bring the HST data from the ICRF3 frame to the VLBI data's frame.}
\label{tab:astrometry}

\begin{tabular}{llcll}
\hline
(1) & (2) & (3) & (4) & (5)\\
Days post merger & Telescope   & Position & Statistical uncertainty & Systematic uncertainty \\ \hline
8               & HST         & 13:09:48.068460 $-$23:22:53.3908         & (0.32, 0.19) & (0.18, 0.34)                                             \\
75              & HSA         & 13:09:48.068634 $-$23:22:53.3908          & (0.14, 0.5)                         &     (0.14, 0.49)                                        \\ 
207             & global-VLBI &   13:09:48.068793 $-$23:22:53.3903       &   (0.19, 0.33)                      &     (0.14, 0.49)                                        \\ 
230             & HSA         &  13:09:48.068835 $-$23:22:53.3908        &           (0.14, 0.4)              &       (0.14, 0.49)                                      \\ 
\end{tabular}
\end{table}

\begin{comment}

\begin{figure}[h]
\plotone{M18_free_astrometry.pdf}
\caption{ Fits of the afterglow models originally analysed in \citet{Mooley18}, obtained using the Bayesian model-fitting framework presented in Section \ref{methods}. The black markers correspond to the image-plane 2-D Gaussian-fit positions of GW170817 from HSA VLBI data (74.7 and 230.25 days post-merger). The errorbars consist of the statistical uncertainties of the image-plane fit (as returned by the \textsc{AIPS} task \textsc{jmfit}) and the astrometric systematic uncertainties of the VLBI data as reported in \citet{Mooley18} and \citet{Mooley22}. The radio counterpart is moving towards the East (or left). The model data points (denoted by crosses) correspond to the flux centroids of the model images, having applied the posterior of each fit parameter to the model. The right-most model data points denote the location of the merger according to the fitted model. The three model data points are connected by lines for each model, to help guide the eye. All data are plotted in the frame of the HSA VLBI observations and are relative to the VLBI phase center, denoted by a cross and intersecting vertical and horizontal dotted lines.} 
\label{fig:freefit}
\end{figure}
\end{comment}

\begin{table}[]
\setlength\tabcolsep{4pt}\makegapedcells
\begin{tabular}{lll}
\hspace{-4em}\makecell[lc]{Fit\\ parameter}          & \makecell[cc]{Description}                                                                                                                        & Prior                       \\ \hhline{===}
$F_0$                  & a global flux scale parameter in Jy units for all VLBI visibility datasets                                                         &    $\propto \left\{
                \begin{array}{ll}
                  1 \quad 5.5\times10^{-6} \le F_0 < 5.5\times10^{-4}\\
                  0 \quad {\rm otherwise}
                \end{array}\right.$
               \\
$\delta_{F,i}$             & \hspace{-4em}\makecell[l]{nuisance parameter accounting for the fractional uncertainty on $F_0$ in\\ the $i^{th}$ VLBI epoch}                                                             & $\propto \text{Gaussian}(\sigma=0.1, \mu=0)$                           \\
PA                     & the orientation of the model images, East of North in degrees                                                                      &                                   $\propto\left\{
                \begin{array}{ll}
                  1 \quad 0 \le \text{PA} < 180\\
                  0 \quad {\rm otherwise}
                \end{array}\right.$                          \\
\ra                     & \hspace{-7.2em}\multirowcell{2}{\makecell[lc]{$^{a}$ the right ascension and declination of the merger location,\\ expressed as an offset from the HSA observations’\\ phase center in milliarcseconds}} &
$\propto \text{Gaussian}(\sigma=0.18, \mu=-7.44)$                               \\
\dec                    &                              &     
$\propto \text{Gaussian}(\sigma=0.34, \mu=0.81)$                        \\
$\delta_{\text{ra,i}}$ & \hspace{-7.2em}\multirowcell{2}{\makecell[lc]{nuisance parameters corresponding to the systematic astrometric\\ uncertainty in RA and Dec associated with a given VLBI epoch $i$ (in mas)}}   &   
$\propto \text{Gaussian}(\sigma=0.14, \mu=0)$ 
\\
$\delta_{\text{dec,i}}$            &        &   $\propto \text{Gaussian}(\sigma=0.49, \mu=0)$                           \\
%$\delta_{\text{ra,2}}$            & \hspace{-7.2em}\multirowcell{2}{\makecell[lc]{Same as above but for the day-230.25 HSA data}} & same as  $\delta_{\text{ra,1}}$                            \\
%$\delta_{\text{dec,2}}$           &      &    same as $\delta_{\text{dec,1}}$                         \\ 
%$\delta_{\text{ra,3}}$            &  \hspace{-7.2em}\multirowcell{2}{\makecell[lc]{Same as above but for the day-206.5 global-VLBI data}}                                                                                                                            &   same as $\delta_{\text{ra,1}}$                          \\
%$\delta_{\text{dec,3}}$           &                                                                                                                                &             same as $\delta_{\text{dec,1}}$                \\
\ratio  & $^b$The ratio of jet viewing angle to jet opening angle at \tp                                                          & 
$\propto \text{Gaussian}(\sigma=2.5, \mu=7.5)$\\
\diff   & \hspace{-3.5em}\makecell[l]{The difference between jet viewing angle\\ and jet opening angle at \tp, in radians}                                   &  
                $\propto \left\{
                \begin{array}{ll}
                  1 \quad 0.05 \le \text{\diff}  < 0.9\\
                  0 \quad {\rm otherwise}
                \end{array}\right.$                            \\
\tp     & $^c$The time in days post-merger at which the afterglow light-curve peaks                                                              & 
$\propto \text{Gaussian}(\sigma=4, \mu=155)$\\

\\ 
$d_L$                  & $^d$The luminosity distance to GW170817 (Mpc)                                                                                         &       
$\propto d_L^2,\, d_{L,\text{min}}=20, d_{L,\text{max}}=50$

\\
$cz$                   & Observed heliocentric group redshift (km\,s$^{-1}$)                                                                                               &      

  $\propto \left\{
                \begin{array}{ll}
                  1 \quad 2000 \le cz  < 4000\\
                  0 \quad {\rm otherwise}
                \end{array}\right.$

\\
$H_0$                  & The Hubble constant (km\,s$^{-1}$\,Mpc$^{-1}$)                                                                                                               &  $\propto \frac{1}{H_0},\, H_{0,\text{min}}=20, H_{0,\text{max}}=120$

\\ \hhline{===}                         
\end{tabular}
\\
$^a$ Here, $\mu$ is the HST position reported in \citet{Mooley22}, in the HSA frame. The $1\sigma$ values correspond to the uncertainty in RA and Dec in the position of the common phase reference source (J1312--2350) that is used to place the HST data from the ICRF3 frame into the HSA data's frame. Where the HST data is not included in the fit, \ra\ and \dec\ have priors that are uniform between $-50$ and 0, and $-50$ and 5 milliarcseconds, respectively. The phase centre of the HSA data is 13:09:48.069 $-23$:22:53.39.  \\

$^b$ Here, the prior is informed by the afterglow light-curve analysis from \citet{Makhathini21}, which directly constrains \ratio. Where fits in our analysis do not include light-curve information, an uninformative prior between 1 and 25 has been used.\\

$^c$ Here, the prior is informed by the afterglow light-curve analysis from \citet{Makhathini21}, which directly constrains \tp. Where fits in our analysis do not include light-curve information, \tp\ is set to 155 days.\\

$^d$ We have verified that our inference is insensitive to these luminosity distance prior bounds by widening the bounds to 1--75\,Mpc to find statistically indistinguishable results to those reported in Section \ref{res:h0}. Indeed, the posterior samples fall within the stricter bounds, as reflected in Figure \ref{fig:corner}. We have kept the narrower bounds purely for computational efficiency.

\caption{Description of Bayesian fit parameters and their priors. A condition is applied to the \ratio\ and \diff\ priors such that the combination of drawn values obeys $0<$\obs$<0.75$\,rad. This upper bound is motivated by the jet model, which is presented in \citet{govreen-segal23} and is calibrated using simulations with \obs\ up to $0.75$ radians.}

%\obs($=\frac{(\theta_{\text{v}}-\theta_{\text{cp}})\frac{\theta_{\text{v}}}{\theta_{\text{cp}}}}{\frac{\theta_{\text{v}}}{\theta_{\text{cp}}}-1})$
 \label{tab:params}
\end{table}

\begin{comment}
    
\begin{figure}[h]
\centering
%\plotone{GW_KDE_corner_pub.png}
\includegraphics[width=\textwidth]{gc1_r2.png}
\caption{Example corner plot from the most comprehensive fit performed in this study, where cosmology is included as described in Section \ref{meth:h0}. This corner plot corresponds to the fit where group catalog 763 and peculiar velocity reconstruction \texttt{TwoMRS\_redshift}, which resulted in the largest evidence of all such peculiar velocity correction combinations. \textbf{KG to polish this plot for publication}} 
\label{fig:full_corner}
\end{figure}
\end{comment}

\begin{table}[]
\setlength\tabcolsep{30pt}\makegapedcells
\begin{tabular}{lll}
\hspace{-4em}\makecell[ll]{Fit parameter}  & \hspace{-4em}\makecell[ll]{Bayesian model average\\mean value} & \hspace{-4em}\makecell[ll]{Bayesian model average\\68\% credible interval} 
\\ \hhline{===}
$F_0$ (Jy)                  & $5.074\times10^{-5}$ & $[4.573\times10^{-5},5.582\times10^{-5}]$ \\
$\delta_{F,1}$             & 0.003 & $[-0.081,0.089]$                           \\
$\delta_{F,2}$             & -0.033 & $[-0.122, 0.057]$                           \\
$\delta_{F,3}$             & 0.024 & $[-0.067,0.114]$                           \\
PA    (degrees)                 & 87.7   & $[80.2, 95.4]$   \\
\ra\ (mas)                     &  $-7.28$  & $[-7.40, -7.16]$ \\
\dec\ (mas)                   &  $-0.82$  & $[-0.98, -0.66]$ \\
$\delta_{\text{ra,1}}$ (mas) &  $0.2$  & $[-0.1, 0.3]$ \\
$\delta_{\text{dec,1}}$ (mas) &  $-0.14$  & $[-0.50, 0.22]$ \\
$\delta_{\text{ra,2}}$ (mas) &  $-0.02$  & $[-0.15, 0.10]$ \\ 
$\delta_{\text{dec,2}}$ (mas) &  $0.01$  & $[-0.44, 0.46]$ \\
$\delta_{\text{ra,3}}$ (mas) &  $-0.06$  & $[-0.18, 0.06]$ \\   
$\delta_{\text{dec,3}}$ (mas) & $0.06$ & $[-0.35, 0.48]$       \\
\ratio & 6.4 & $[4.3,8.5]$ \\
\diff\ (radians)  & 0.260  & $[0.241, 0.278]$ \\
\tp\   (days)  & 154.7 & $[150.8,158.7]$ \\
$d_L$ (Mpc) & 44.0  & $[42.5,45.6]$ \\
$cz$ (km\,s$^{-1}$) & 3002 & $[2938,3061]$ \\
$H_0$ (\kms) & $65.5$ & $[61.1, 69.9]$ \\
\hhline{===}                         
\end{tabular}
\\

\caption{Bayesian model average posterior means and 68\% credible intervals. These are obtained by combining all posterior samples from the 28 fits corresponding to different peculiar velocity corrections described in Section \ref{meth:h0} and presented in Section \ref{res:h0}, and weighting them by their respective Bayes evidence. The fit parameters and their priors are described in Table \ref{tab:params}.}

%\obs($=\frac{(\theta_{\text{v}}-\theta_{\text{cp}})\frac{\theta_{\text{v}}}{\theta_{\text{cp}}}}{\frac{\theta_{\text{v}}}{\theta_{\text{cp}}}-1})$
 \label{tab:posterior}
\end{table}

\section{Independent reduction of Global-VLBI data}
\label{reduction}
We obtained the correlated data corresponding to project code GG084a from the European VLBI Network (EVN) archive and followed the same initial steps described in G19 up to and including ionospheric delay corrections through the \textsc{tecor} AIPS task. Given the length of the observing run, there is no time at which all antennas are observing simultaneously. Therefore, we split the calibration into three time-windows and corresponding groups of antennas, each with their own reference antenna. For each group, we corrected for instrumental delays and phase offsets between frequency subbands (IFs) with \textsc{fring} using data from a single scan from the respective time-window. The resulting three SN tables were applied to the corresponding groups of antennas with \textsc{clcal} using reference antennas consistent with those that were used with \textsc{fring}. Global fringe fits were performed with \textsc{fring} on phase calibrator source J1311$-$2329, using all available antennas in each time-window and combining all 16 IFs to increase signal-to-noise. The resulting solutions were clipped for delay and rate outliers and were applied to J1311$-$2329, the target and the check-source (J1312$-$2350) at the corresponding time windows with \textsc{clcal}. The resulting calibrated phase calibrator source was averaged in frequency to a single 16\,MHz channel per IF with \textsc{split} and saved as a \textsc{uvfits} file with \textsc{fittp}. The calibrated J1311$-$2329 data was read into \textsc{difmap} where a model of the source was created using phase and amplitude self-calibration up to 0.5 and 20 minutes, respectively. The resulting model was read back into AIPS and the global \textsc{fring} steps above were repeated using this model. Each resulting SN table was used in a round of amplitude calibration on a 20-minute timescale with \textsc{calib} on the unaveraged J1311$-$2329 data. The resulting solutions were applied to the target and check-source with \textsc{clcal} and \textsc{split} was used to produce uvfits data for each source. The target is offset from the phase center by approximately 280\,mas in RA and 40\,mas in Dec. An offset of such a magnitude would result in significant smearing should the data undergo time and frequency averaging. Hence, the phase center of the calibrated target visibility data was shifted using \textsc{uvfix} before averaging in frequency to a single 16\,MHz channel per IF and in time by the scan length (150 seconds). The data was saved as a \textsc{uvfits} file with \textsc{fittp}. Imaging and minor station-based flagging of the calibrated target visibilities were conducted in \textsc{difmap} \citep{difmap}. We fit the visibilities with a single circular Gaussian component to produce a clean image with natural weighting (raised using a powerlaw index of $-1.5$ to give optimal image results). The size of the synthesised beam is $3.2 \times 6.1$\,mas (minor and major axes) with a PA of $-5$\degree. The peak flux in the clean image is  47\,\ujy\ and the rms is approximately 11\,\ujy. The target is detected in the clean image by the AIPS task \textsc{jmfit}, which fits a Gaussian in the image-plane, with a signal-to-noise (S/N) value of 4.3. The fitted Gaussian's deconvolved size is consistent with zero, implying that the source is unresolved in the image plane. The \textsc{jmfit} position is RA $= 13^{\text{h}} 09^{\text{m}} 48^{\text{s}}.068752$, Dec $=-23^{\text{d}}22^{\text{m}}53^{\text{s}}.3915$. As a check, we also ran \textsc{jmfit} on the published image of GW170817 from G19 (which had a synthesized beam of $2.9\times7.0$\,mas, PA $=-6\degree$) which yielded a consistent S/N of 4.1, and similarly indicated that the source is unresolved .\footnote{\url{https://github.com/omsharansalafia/radiogw17.git}}
%G19 beamsize is 2.9x7.0 mas

%in actuality for the imaging the 170817 data was averaged in time by only 10 seconds. We average over the course of a full scan only for the modelfitting. But I don't think this distinction is important... can check by imaging the scan ave data to see if makes a difference (Need to fittp that data from AIPS).

We calculate the astrometric shifts required to place the global-VLBI data into the frame of the HSA data by comparing the position of check-source J1312$-$2350. We used the AIPS task \textsc{UVSUB} to divide the global-VLBI target data by the model that was obtained for J1312$-$2350 by \citet[]{Mooley18}. A perfect match between model and data would result in the divided dataset showing a point source at the phase center of the resultant image, while any astrometric offset between the model and the global-VLBI data would manifest as an offset from the resulting image's phase center. As such, the divided uvfits data was imaged in difmap using a single point source model component and the resulting clean image was fitted using \textsc{jmfit} which reported an astrometric offset of $-0.562$ mas in RA and $-1.204$ mas in Dec. These values are used to shift the global-VLBI data into the HSA frame when including those data into our Bayesian fit by folding them into that epoch's prior position. The position of GW170817 at day 207 in the global-VLBI data in the HSA data frame is then RA $= 13^{\text{h}}09^{\text{m}}48^{\text{s}}.068793$, Dec $= -23^{\text{d}}22^{\text{m}}53^{\text{s}}.3903$. As astrometric systematic uncertainty on this dataset, M22 use 0.035 mas in RA and 0.14 mas based on simulations from \citet{pradel06}. However, as the simulations of \citet{pradel06} do not incorporate ionospheric effects, which are likely to be comparable to or larger than the modeled effects at the observing frequency of 4.5 GHz, we prefer a more conservative positional uncertainty estimation. Given the comparable baseline lengths, observing frequency and calibrator, we instead apply the same uncertainties that we use for the HSA data so that $\delta_{\text{ra,3}}=0.14$\,mas and $\delta_{\text{dec,3}}=0.49$\,mas. Lastly, we compare our day-207 position with that which was previously reported by G19 and shifted into a common astrometric frame in M22. When in a common frame, we find that the difference in position between our day-207 reduction and that reported in M22 is 0.46\,mas in RA and 0.48\,mas in Dec; this offset is within $2\sigma$ in RA and $1\sigma$ in Dec of the  total (statistical and systematic) uncertainties as estimated by both M22 and this work. The astrometric difference can be visualized in Figure \ref{fig:scaling}.

\section{The implied priors on jet geometry}
\label{app:prior}

\begin{figure}[h]
\centering
%\plotone{GW_KDE_corner_pub.png}
\includegraphics[width=0.6\textwidth]{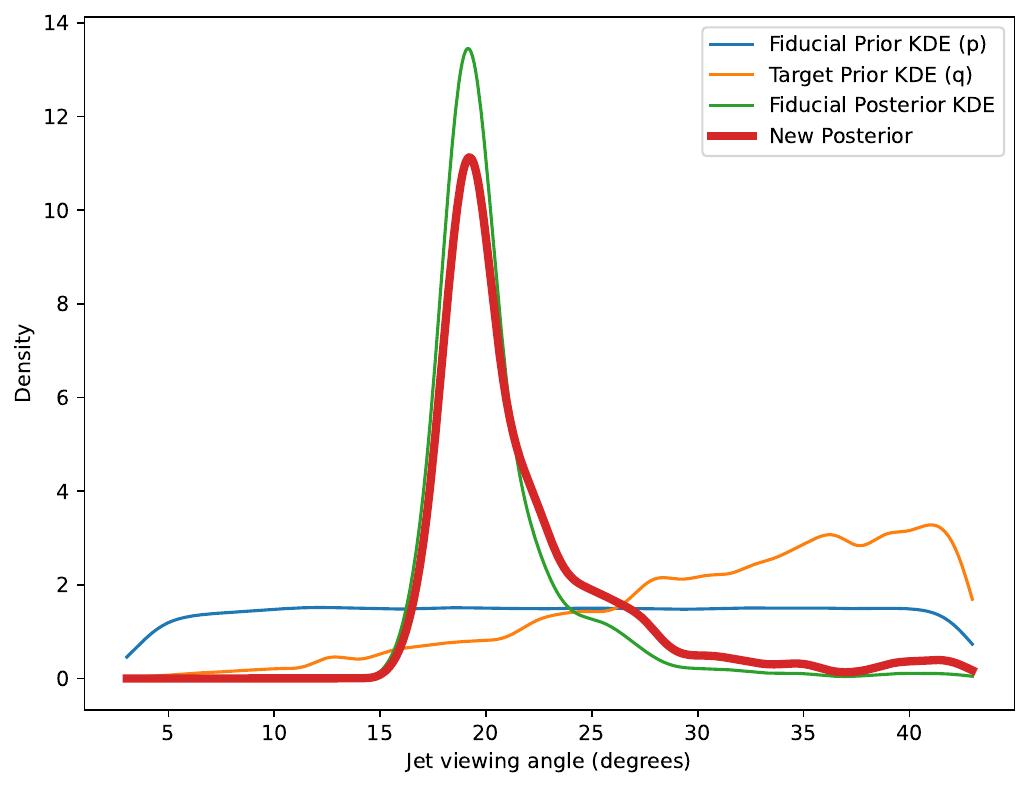}
\caption{Comparison of the viewing angle KDEs described in Appendix \ref{app:prior} associated with the 4-epoch fit (where uninformative priors on jet geometry are desired): the implied fiducial prior actually used in this study ($p$, blue; uniform in \obs), the resulting fiducial posterior (green), the desired target prior ($q$, orange; uniform in $\cos(\text{\obs})$, subject to the constraint \cp$<$\obs) and the `new' corrected posterior that results from scaling the fiducial posterior by $\frac{q}{p}$.} 
\label{fig:kdes}
\end{figure}

In Section \ref{res:scaling} and Figure \ref{fig:geometries}, viewing angle credible intervals are reported based on the fits that are described in Section \ref{meth:scaling}. Where we do not wish to employ light-curve information in our prior beliefs about the jet geometry, we would seek to apply prior distributions that are uniform in both $\cos(\text{\obs})$ (isotropic viewing angles) and opening angle \cp, subject to the constraint that \cp$<$\obs (a requirement of our jet model). However, as described in Section 
\ref{meth:scaling}, we cannot directly fit for \obs\ and \cp, but rather the two combinations of those parameters: \ratio\ and \diff. In our analysis, \obs\ credible intervals are reported by algebraically rearranging \ratio\ and \diff\ posterior samples into \obs\ posterior samples. Accordingly, in the `no light-curve information' fits, uninformative priors on both \diff\ and \ratio\ are used (see Table \ref{tab:params}). However, the \textit{implied} fiducial priors on \cp\ and \obs\ would then actually be inconsistent with the desired `target' distributions described above. Here, we wish to assess the impact of the discrepancy between fiducial and target prior on the fiducial posteriors that are reported in this study. We do so by fitting KDEs to the fiducial priors ($p$), target priors ($q$), and fiducial posteriors from the four-epoch fit (bottom left corner of Figure \ref{fig:geometries}). The corrected posterior is obtained by multiplying the fiducial posterior KDE distribution by $\frac{q}{p}$. A comparison of this new posterior to the other distributions is shown in Figure \ref{fig:kdes}. We find that the 68\% credible interval of the new \obs\ posterior is about 3 degrees broader than the fiducial KDE (18.2--25.3 degrees, c.f. 18.0--22.4 degrees), demonstrating that the implied priors act to push the posterior to smaller viewing angles. We emphasize that where informative, Gaussian, priors on \ratio\ are used in our fits (based on posteriors from independent analyses of the afterglow light-curve), no such corrections are required; the consideration discussed here is strictly relevant to the case where uninformative jet geometry priors are desired. 

%we use an uninformative prior unless the \ratio\ posterior from independent afterglow light-curve analyses is employed as a prior. 

%position of J1312 in my selfcal version: '13h12m48.75802993s -23d50m46.954381s'. This is offset from Ghirlanda by 0.1 in RA and -0.475 in Dec.
%The calibrated data was averaged in time by 10 seconds and in frequency to a single 16-MHz channel per IF.

\bibliography{references}{}
\bibliographystyle{aasjournal}

\end{document}